\documentclass{aa}  

\usepackage{graphicx}
\usepackage{txfonts}
\usepackage{lipsum}
\usepackage{subcaption}         
\usepackage{lscape}             
\usepackage{placeins}           
\usepackage{CJKutf8}
\usepackage[T1]{fontenc}
\usepackage[colorlinks=true, linkcolor=blue, citecolor=blue, filecolor=blue, urlcolor=blue]{hyperref}
\usepackage{siunitx}
\usepackage{bm}

\begin{document} 
    \title{Late-infall-induced formation of giant planets, multigenerational planetesimals, and disk substructures}

\author{
Haichen Zhao
\begin{CJK*}{UTF8}{gkai}(赵海辰)\end{CJK*}
\inst{1}
\and
Tommy Chi Ho Lau
\begin{CJK*}{UTF8}{bkai}(劉智昊)\end{CJK*}
\inst{2}
\and
Joanna Dr{\k{a}}{\.z}kowska
\inst{1}
\and
Tilman Birnstiel
\inst{3,4}
\and
Sebastian M. Stammler
\inst{3}
}

   \institute{Max Planck Institute for Solar System Research, Justus-von-Liebig-Weg 3, 37077 Göttingen, Germany\\
   \email{zhaoh@mps.mpg.de}
   \and
   Department of Astronomy and Astrophysics, University of Chicago, Chicago, IL 60657, USA
   \and
   University Observatory, Faculty of Physics, Ludwig-Maximilians-Universit\"at M\"unchen, Scheinerstr. 1, D-81679 Munich, Germany
   \and
   Exzellenzcluster ORIGINS, Boltzmannstr. 2, D-85748 Garching, Germany
   }

   \date{Received 10 June 2026; accepted 17 July 2026}

  \abstract
    {Late infall can replenish the building materials of planets in protoplanetary disks and dramatically alter their structural evolution. The resulting pressure bumps effectively accumulate dust, facilitate grain coagulation, and trigger planetesimal formation via the streaming instability. In this work, we investigated the potential for planetesimal and planet formation, as well as the emergence of observable substructures, in disks undergoing late-stage infall. We utilized a comprehensive modeling framework that couples dust coagulation and dynamics, planetesimal formation, $N$-body gravity, planetary growth, and planet-disk interactions. Our results show that the abundant dust supply and the migration barrier created by the infalling gas enable the rapid formation of gas giants via pebble and gas accretion within one million years, even at large orbital distances ($\sim 70$ au). These giants, in turn, exert torques that generate multiple secondary disk substructures, fostering multigenerational planetesimal formation and resulting in diverse planetary system configurations. The planetesimals exhibit distinct dynamical properties that are determined by their formation epoch and environment, which are analogous to the small-body populations in the outer Solar System. Both the infall- and planet-induced substructures are clearly visible in synthetic 1.3-mm continuum observations, closely resembling the multi-ring disks detected in ALMA surveys. Our model provides a new perspective on the origin of distant giant planets, long-lasting planetesimal formation, and the prevalence of disks with multiple substructures.}

   \authorrunning{Haichen Zhao et al.}
   \maketitle
   \nolinenumbers

\section{Introduction} \label{sec:intro}

For decades, protoplanetary disks were primarily modeled as isolated systems. Recently, however, the influence of the external environment on disk evolution has emerged as a pivotal factor in shaping the process of planet formation. These environmental interactions can be destructive, such as external photoevaporation driven by nearby massive stars \citep{Winter22}, or constructive, such as the late-stage infall of material from the surrounding interstellar medium \citep{Pineda23}. Late infall refers to the accretion of gas and dust onto a disk after the initial collapse of the parent molecular cloud core has largely subsided, typically during the late Class I or early Class II phases. This process is driven by two primary mechanisms: the capture of cloudlets during close encounters with low-mass cloud fragments \citep{Dullemond19}, and Bondi-Hoyle-Lyttleton accretion as the disk traverses a low-density interstellar medium \citep{Krumholz06}.

A wealth of observations has identified numerous protostellar systems undergoing late infall, with examples spanning molecular line emission \citep[e.g.,][]{Valdivia-Mena22,Gupta26}, continuum emission \citep[e.g.,][]{Cacciapuoti24}, and scattered light \citep[e.g.,][]{Ginski21}. These systems are characterized by streamers that transport material from the large-scale environment to the disk. Both theoretical models and recent surveys suggest that such events are remarkably common. For instance, a census by \cite{Garufi24} of Class II objects in the Taurus star-forming region revealed that approximately one-third of the sample exhibits evidence of environmental interaction. Furthermore, \cite{Winter24} used an excursion set formalism to argue that 20--70\% of protoplanetary disks may consist predominantly of material accreted during the latter half of their lifetimes.

This influx of material offers a compelling solution to the ``mass budget problem,'' where the measured dust masses in Class II disks often appear insufficient to account for the observed exoplanet population \citep{Manara18}. While the early onset of planet formation is one potential explanation \citep{Drazkowska18}, late infall provides a mechanism to replenish the disk and enable planet formation at later epochs. The detection of dust within an infall streamer by \cite{Cacciapuoti24} confirms that these events supply the necessary solids alongside gas. Beyond simple mass replenishment, hydrodynamic simulations show that late infall fundamentally alters disk dynamics by inducing substructures such as rings, vortices \citep{Bae15, Kuznetsova22}, and spirals \citep{Calcino25, Huhn26}. While observations have confirmed the coexistence of infall streamers and disk substructures \citep[e.g.,][]{Yen19, Ginski21}, the precise dynamical link between them remains a subject of active investigation.

The journey from sub-micron grains to gas giants is a multi-stage process spanning roughly 40 orders of magnitude in mass \citep[see][for a review]{Drazkowska23}. Two major bottlenecks hinder this growth: the fragmentation and drift barrier, which limits dust growth to centimeter-to-meter-sized particles \citep{Blum08}, and the low efficiency of planetesimal accretion in the outer disk \citep{Pollack96}. The first can be bypassed via the streaming instability, where dynamical interaction between gas and dust concentrates solids into dense clumps that subsequently undergo gravitational collapse into planetesimals (\citealp{Youdin05, Johansen07}; see \citealp{Simon24}, for a review). The second is mitigated by pebble accretion, in which centimeter-to-meter-sized solids are captured by a protoplanet’s gravity, aided by gas drag (\citealp{Ormel10, Lambrechts12}; see \citealp{Johansen17}, for a review). Both mechanisms benefit from the high dust concentrations and increased pebble flux found within infall-induced substructures, making them ideal sites for planetesimal formation and rapid planetary growth.

As protoplanets grow to a few tens of Earth masses, their gravitational torques on the disk begins to carve gaps and create secondary pressure bumps at the gap edges. This evolution shifts the disk from being shaped by environmental infall to being governed by internal planetary feedback, suggesting a sequential, multigenerational planet formation process. High-resolution continuum observations by ALMA have identified numerous disks with multiple rings and gaps \citep[e.g.,][]{Andrews18, Long18}. These structures are often interpreted as evidence of massive planets residing within the gaps \citep{Zhang18}, while the surrounding rings may serve as sites for ongoing planetesimal formation \citep{Stammler19}. Such a progression is consistent with meteoritic records of the Solar System, which indicate that Jupiter formed early and was followed by persistent planet formation spanning several million years \citep[see][for a review]{Kleine20}. By integrating dust and gas evolution with various stages of planetary growth, \cite{Lau24} proposed a framework where early-formed giant planets generate new formation sites at larger radii, fostering subsequent generations of planets.

In this study, we utilized the \texttt{DustPy}\footnote{\texttt{DustPy} v1.0.8 was used in this work.} code \citep{dustpy} to model the coupled evolution of gas and dust in a disk where late infall generates an axisymmetric substructure. Our previous work \citep{Zhao25} demonstrated that these infall-induced pressure bumps can successfully concentrate dust to form planetesimal belts. Here, we extend that analysis by coupling \texttt{DustPy} with the $N$-body integrator \texttt{SyMBAp}\footnote{\texttt{SyMBAp} v1.8 was used in this work.} \citep{symbap}, a parallelized version of the Symplectic Massive Body Algorithm \citep[\texttt{SyMBA};][]{symba}. This allows us to investigate the formation of giant planets, their subsequent feedback on the disk, and the resulting potential for multigenerational planetesimal formation. Finally, we produce synthetic observations to identify the observational signatures of these processes.

This paper is organized as follows. Section \ref{sec:method} details our numerical methods and simulation setup. Section \ref{sec:result} presents the results and physical evolution of the modeled systems. Synthetic observations are analyzed in Section \ref{sec:synobs}. We discuss the broader implications in Section \ref{sec:discuss} and summarize our conclusions in Section \ref{sec:conclu}.

\section{Methods} \label{sec:method}

We assumed an axisymmetric protoplanetary disk around a Solar-mass star. The gas component undergoes viscous evolution, while dust particles are subject to transport via advection and diffusion alongside collisional evolution, including coagulation and fragmentation. Planetary growth was modeled through a combination of collisions, pebble accretion, and gas accretion. Furthermore, planet-disk interactions drive both gap opening and orbital migration. The disk model, the implementation of late infall, and the criteria for planetesimal formation followed \citet{Zhao25}. The numerical realization of planetesimals, as well as the prescriptions for planetary growth and planet-disk interactions, are largely consistent with those adopted by \citet{Lau25}, with minor modifications specific to this study.

\subsection{Disk model} \label{ssec:met_disk}

\subsubsection{Gas component}  \label{sssec:met_disk_gas}

The initial radial profile of the gas surface density, prior to the onset of infall, follows the self-similar solution from \citet{Lynden-Bell74}:
\begin{equation}
    \Sigma_{\rm g,init} = \frac{M_{\rm disk}}{2\pi r_{\rm c}^2} \left(\frac{r}{r_{\rm c}} \right)^{-1} \exp \left(-\frac{r}{r_{\rm c}} \right),
\label{eq:Sig_init}
\end{equation}
where $r$ is the radial distance to the star, $M_{\rm disk} = 0.01\, M_{\odot}$ represents the initial disk mass, and $r_{\rm c} = 200$ au is the characteristic radius.

We assume a passively irradiated disk in vertical hydrostatic equilibrium, yielding the following radial profiles for midplane temperature and the disk aspect ratio:
\begin{align}
    &T \approx 221.3 \left(\frac{r}{\rm au} \right)^{-1/2}\,{\rm K}, \\
    &\hat{h}_{\rm g} \equiv \frac{H_{\rm g}}{r} \approx 0.03 \left(\frac{r}{\rm au} \right)^{1/4},
\end{align}
where $H_{\rm g}$ is the gas scale height. The gas volume density, pressure, and the dimensionless pressure gradient parameter at the midplane are calculated as
\begin{align}
    &\rho_{\rm g,mid} = \frac{\Sigma_{\rm g}}{\sqrt{2\pi}H_{\rm g}}, \\
    &P = \rho_{\rm g,mid}c_{\rm s}^2, \\
    &\eta = -\frac{\hat{h}_{\rm g}^2}{2} \frac{\partial \ln{P}}{\partial \ln{r}},
\end{align}
respectively.

The gas evolution is governed by the viscous advection-diffusion equation \citep{Luest52, Lynden-Bell74}, 
\begin{equation}
    \frac{\partial \Sigma_{\rm g}}{\partial t} = \frac{3}{r} \frac{\partial}{\partial r} \left[\sqrt{r} \frac{\partial}{\partial r} \left(\nu \Sigma_{\rm g}\sqrt{r} \right) - \frac{2\Lambda \Sigma_{\rm g}}{3\Omega_{\rm K}}\right] + S_{\rm g,infall},
    \label{eq:adv-diff}
\end{equation}
where $\nu$ is the kinematic viscosity, and $\Lambda$ is the injection rate of the specific angular momentum contributed by potentially forming massive planets (see Section \ref{sssec:met_itact_gap}). The external source term representing late infall,
\begin{equation}
    S_{\rm g,infall} = \frac{\dot{M}_{\rm infall}}{A_{\rm norm}} \exp\left[-\frac{\left(r - r_{\rm infall}\right)^2}{2 \sigma_r^2}\right],
    \label{eq:infall}
\end{equation}
assumes that infalling gas follows an axisymmetric Gaussian distribution centered at $r_{\rm infall} = 80$ au with a radial width defined by the standard deviation $\sigma_r = 15\% \times r_{\rm infall}$. The normalization factor $A_{\rm norm}$ ensures that the integral of $S_{\rm g,infall}$ over the disk area equals the total infall rate $\dot{M}_{\rm infall} = 1.2 \times 10^{-7}\,M_{\odot}\,{\rm yr}^{-1}$. Infall begins at the start of the simulation and persists for $5 \times 10^4$ years.

We assumed the specific angular momentum of the infalling material matched that of the local disk at each radius. While this approach does not capture the complex 3D gas dynamics or angular momentum mismatches investigated in some hydrodynamic studies, this axisymmetric approximation provides a robust and computationally efficient framework for examining the radial transport and growth of dust. Turbulent transport is modeled using the \citet{Shakura73} $\alpha$ parameterization for kinematic viscosity:
\begin{equation}
    \nu = \alpha c_{\rm s} H_{\rm g},
    \label{eq:alpha}
\end{equation}
where $\alpha = 10^{-4}$ and $c_{\rm s}$ is the isothermal sound speed. We neglected the dynamical back-reaction of dust onto the gas.

\subsubsection{Dust component}  \label{sssec:met_disk_dust}

The initial disk and the infalling material are both assumed to have a dust-to-gas ratio consistent with Solar metallicity, $Z_{\rm init} = Z_{\rm infall} = 0.01$. The initial dust size distribution follows the Mathis-Rumpl-Nordsieck (MRN) power law for the interstellar medium \citep{MRN77}, $n(a) \propto a^{-7/2}$, ranging from a minimum size of \SI{0.1}{\micro\meter} to a maximum of \SI{1}{\micro\meter}. We assumed a bulk material density of $1.4 \,\rm g\,cm^{-3}$.

\texttt{DustPy} solved for dust evolution using a mass grid-based model. The Stokes number, ${\rm St}$, of the particles was calculated across both the Epstein and Stokes I regimes. The dust scale height for each mass bin $i$ is determined following \citet{Dubrulle95}:
\begin{equation}
    H_{\text{d},i} = H_{\rm g} \sqrt{\frac{\alpha}{\alpha+\text{St}_i}},
    \label{eq:H_d}
\end{equation}
which allows for the calculation of the midplane dust volume density:
\begin{equation}
    \rho_{\text{d,mid},i} = \frac{\Sigma_{\text{d},i}}{\sqrt{2\pi}H_{\text{d},i}}.
\end{equation}

The radial transport of solids is governed by the advection-diffusion equation \citep{Clarke88}, to which we have added an external source term to account for the infalling dust. Collisional evolution is determined by solving the Smoluchowski equation \citep{Smoluchowski1916}, including the effects of coagulation and fragmentation. We set the fragmentation threshold to $v_{\rm frag} = 350\,\rm cm\,s^{-1}$, where relative velocities exceeding this value result in destructive outcomes. For a comprehensive description of the \texttt{DustPy} algorithms, we refer the reader to \citet{dustpy}.

\subsection{Planetesimal formation} \label{ssec:met_pltform}

The criteria for planetesimal formation are determined by two processes that occur on sub-grid scales and are thus parameterized in our model. First, the streaming instability (SI) is triggered when the dust-to-gas density ratio at the disk midplane reaches unity:
\begin{equation}
    \epsilon_{\rm mid} = \frac{\rho_{\rm d,mid}(10^{-3} < \text{St} < 1)}{\rho_{\rm g,mid}} \geq 1.
    \label{eq:eps_mid}
\end{equation}
As suggested by several studies (e.g., \citealp{Carrera15}; \citealp{Yang17}; see \citealp{Simon24}, for review), only moderately coupled particles ($10^{-3} < {\rm St} < 1$) are included in this calculation.

Second, the gravitational collapse of the dense dust filaments produced by the SI is governed by a Toomre-like instability parameter derived by \citet{Klahr21}:
\begin{equation}
    Q_{\rm p} = \sqrt{\frac{\delta}{\text{St}_{\rm avg}}} \frac{c_{\rm s} \Omega_{\rm K}}{\pi G \Sigma_{\rm d,local}} < 1,
    \label{eq:Qp}
\end{equation}
where ${\rm St}_{\rm avg}$ is the mass-averaged Stokes number of the pebbles participating in the SI, and $\Omega_{\rm K}$ is the Keplerian frequency. The small-scale particle diffusion parameter, $\delta$, is set to $10^{-6}$, consistent with the SI simulations of \citet{Schreiber18}. The local dust surface density within a filament, $\Sigma_{\rm d,local}$, is assumed to be $10 \times \Sigma_{\rm d}(10^{-3} < {\rm St} < 1)$, reflecting the fact that filaments in SI simulations are typically an order of magnitude denser than the background dust density prior to collapse \citep[e.g.,][]{Simon16, Schaefer17}.

When both criteria (Eqs. \ref{eq:eps_mid} and \ref{eq:Qp}) are met within a radial cell, dust species $i$ with $10^{-3} < {\rm St}_i < 1$ were converted into planetesimals following the prescriptions of \citet{Drazkowska16} and \citet{Schoonenberg18}:
\begin{equation}
    \frac{\partial \Sigma_{\rm plts}}{\partial t} = \sum\limits_{10^{-3} < \text{St}_i < 1} \mathcal{P}_{\rm pf} \zeta \Sigma_{\text{d},i} \text{St}_i \Omega_{\rm K}.
\end{equation}
Here, $\mathcal{P}_{\rm pf}$ is a smooth activation function motivated by \citet{Miller21}:
\begin{equation}
    \mathcal{P}_{\rm pf} = \frac{1}{2} \left[1 - \tanh \left(\frac{Q_{\rm p} - 0.75}{0.2}\right) \right].
    \label{eq:act}
\end{equation}
We adopted a conservative planetesimal formation efficiency of $\zeta = 10^{-6}$.

The resulting planetesimal surface density, $\Sigma_{\rm plts}$, serves as a probability density function for determining the radial locations of $N$-body particles. The mass of each newly formed planetesimal was drawn from the initial mass function (IMF) derived by \citet{Gerbig23}, in which the probability density is proportional to the instability growth rates at different scales. The maximum and minimum masses, corresponding to the largest and smallest unstable scales, are
\begin{equation}
    m_{\rm max,min} = \frac{9}{8}\sqrt{\frac{\pi}{2}} \left(\frac{\delta}{\rm St_{avg}}\right)^{3/2} \hat{h}_{\rm g}^3 \left(\frac{1}{Q_{\rm p}} \pm \sqrt{\frac{1}{Q_{\rm p}^2} - 1} \right)^2 M_{\odot}.
\end{equation}
The most probable mass, corresponding to the fastest-growing mode, is
\begin{equation}
    m_{\rm fgm} = \frac{9}{8}\sqrt{\frac{\pi}{2}} \left(\frac{\delta}{\rm St_{avg}}\right)^{3/2} \hat{h}_{\rm g}^3 Q_p^2 M_{\odot}.
\end{equation}
All stochastic sampling was performed via inverse transform sampling.

The sampled mass was subtracted from the planetesimal surface density of the nearest radial cells. At each communication interval between \texttt{DustPy} and \texttt{SyMBAp}, this realization process continued until the total remaining mass in the planetesimal surface density is less than the sampled mass. To avoid a bias toward lower masses, we retained the last drawn ${\rm Uniform}[0,1)$ random variate for the first realization of the subsequent step. This approach differs from \citet{Lau25}, who retained the mass itself. Our modification accounts for the fact that planetesimal formation may be intermittent or occur simultaneously in distant regions where the local conditions, and thus the IMF, vary. Initial eccentricities and inclinations were drawn from Rayleigh distributions with scale parameters of $10^{-6}$ and $5 \times 10^{-7}$ radians, respectively, while other orbital elements are drawn uniformly from 0 to $2\pi$.

\subsection{Planetary growth} \label{ssec:met_pltgrow}

\subsubsection{Pebble accretion}  \label{sssec:met_pltgrow_peb}

Collisions between $N$-body particles are assumed to be perfectly inelastic. The total pebble accretion rate is determined by summing the product of the pebble mass flux, $\dot{M}_{{\rm peb},i}$, and the accretion efficiency, $\varepsilon_{{\rm PA},i}$, across all dust species $i$ within the range $10^{-3} < {\rm St}_i < 1$:
\begin{equation}
    \dot{m}_{\rm PA} = \sum_i \varepsilon_{{\rm PA},i} \dot{M}_{\text{peb},i}.
\end{equation}
The pebble mass flux is defined as
\begin{equation}
    \dot{M}_{\text{peb},i} =2\pi r v_{{\rm drift},i} \Sigma_{{\rm d},i},
\end{equation}
where $v_{{\rm drift},i} = 2{\rm St}_i|\eta|r\Omega_{\rm K}$ is the radial drift velocity \citep{Weidenschilling77}. We calculated the pebble accretion efficiency for each $N$-body particle following the prescriptions of \citet{LiuOrmel18} and \citet{OrmelLiu18}. This approach accounts for both the settling and ballistic regimes and combines the 2D and 3D limits. The mass of accreted pebbles was subtracted from the corresponding dust mass bin and radial cell during the subsequent communication step. Note that an explicit pebble isolation mass \citep{Lambrechts+14} is not required in our framework; because dust and gas evolve self-consistently, the gap opened by a sufficiently massive planet (see Section \ref{sssec:met_itact_gap}) naturally halts the pebble flux.

\subsubsection{Gas accretion}  \label{sssec:met_pltgrow_gas}

Gas accretion is modeled as a two-stage process: an initial thermal contraction phase, where hydrostatic equilibrium is maintained, followed by a runaway accretion phase once a critical mass is reached.

In the thermal contraction phase, the accretion rate is limited by the cooling of the gaseous envelope. A significant heat source in this stage is the ongoing pebble accretion, which opposes envelope contraction. We adopted the modified prescription from \citet{Chambers21}, based on \citet{Piso14} and \citet{Bitsch15}, to account for this heating:
\begin{equation}
\begin{split}
\dot{m}_{\text{cont}} = &\max \Bigg[
  0,\ 
  4.375 \times 10^{-9} 
  \left( \frac{\kappa}{\text{cm}^2\,\text{g}^{-1}} \right)^{-1} \\
  &\times\left( \frac{\rho_{\rm c}}{5.5\,\text{g}\,\text{cm}^{-3}} \right)^{-1/6}
  \left( \frac{m_{\rm c}}{M_\oplus} \right)^{11/3}
  \left( \frac{m_{\text{env}}}{M_\oplus} \right)^{-1} \\
  &\times\left( \frac{T}{81\,\text{K}} \right)^{-1/2}
  M_\oplus\,\text{yr}^{-1}
  - 15\,\dot{m}_{\text{PA}}
\Bigg],
\end{split}
\label{eq:kappa}
\end{equation}
where $\kappa = 0.01\,{\rm cm}^2\,{\rm g}^{-1}$ is the envelope opacity, $\rho_{\rm c} = 5.5\,{\rm g}\,{\rm cm}^{-3}$ is the core density, $m_{\rm c}$ is the core mass, and $m_{\text{env}}$ is the envelope mass. 

The runaway gas accretion phase begins when $m_{\text{env}} \geq m_{\rm c}$. Here, the envelope collapses on the Kelvin-Helmholtz contraction timescale \citep{Ikoma00}. Incorporating the pebble accretion heating term from \citet{Chambers21}, the rate is
\begin{align}
    &\dot{m}_{\rm run} = \frac{m_{\rm p}}{\tau_{\rm KH}} - 15\,\dot{m}_{\text{PA}}, \label{eq:KH} \\
    &\tau_{\rm KH} = 10^8 \left(\frac{m_{\rm p}}{M_{\oplus}}\right)^{-2.5} \left( \frac{\kappa}{\text{cm}^2\,\text{g}^{-1}} \right)\,\rm yr,
\end{align}
where $m_{\rm p}$ is the total planetary mass.

Growth is further constrained by the gas supply from the disk. We adopted the limit from \citet{Tanigawa02}:
\begin{equation}
    \dot{m}_{\rm flow} = 0.29\Sigma_{\rm g} r^2 \Omega_{\rm K} \left(\frac{m_{\rm p}}{M_{\odot}}\right)^{4/3}  \hat{h}_{\rm g}^{-2}. \label{eq:gas_flow}
\end{equation}
Finally, we assumed the accretion rate cannot exceed 80\% of the local disk accretion rate \citep{Lubow06}:
\begin{equation}
    \dot{m}_{\rm acc} = 0.8 \dot{M}_{\rm disk} = 0.8\times 2\pi r\Sigma_{\rm g}|v_{\text{g},r}|, \label{eq:gas_acc}
\end{equation}
where $v_{\text{g},r}$ is the radial velocity of the gas. The resulting gas accretion rate, $\dot{m}_{\rm g}$, is summarized as
\begin{equation}
\dot{m}_{\rm g}=
\begin{cases}
\min(\dot{m}_{\text{cont}}, \dot{m}_{\rm flow}, \dot{m}_{\rm acc}), & m_{\text{env}} < m_{\rm c}, \\
\min(\dot{m}_{\rm run}, \dot{m}_{\rm flow}, \dot{m}_{\rm acc}),   & m_{\text{env}} \geq m_{\rm c}.
\end{cases}
\end{equation}

\subsubsection{Physical radius}  \label{sssec:met_pltgrow_rad}

For planetesimals with masses below $0.1\,M_{\oplus}$, we calculated the physical radius assuming an internal density of $\rho_{\rm s} = 1.5\,{\rm g}\,{\rm cm}^{-3}$. For the mass range $0.1\,M_{\oplus} \leq m_{\rm p} < 5\,M_{\oplus}$, we adopted the mass-radius relationship for rocky planets from \citet{Seager07}:
\begin{equation}
\begin{split}
    \log\left(\frac{R_{\rm p}}{3.3\,R_{\oplus}}\right) = &-0.209 + \frac{1}{3}\log\left(\frac{m_{\rm p}}{5.5\,M_{\oplus}}\right)\\
    &- 0.08 \left(\frac{m_{\rm p}}{5.5\,M_{\oplus}}\right)^{0.4}.
\end{split}
\end{equation}
For $m_{\rm p} \geq 5\,M_{\oplus}$, we followed the relationship used in \citet{Matsumura17}:
\begin{equation}
    R_{\rm p} = 1.65\sqrt{\frac{m_{\rm p}}{5\,M_{\oplus}}} R_{\oplus}.
\end{equation}

\subsection{Planet-disk interactions} \label{ssec:met_itact}

\subsubsection{Gap opening} \label{sssec:met_itact_gap}

We assumed that a planet begins to gravitationally perturb the disk when the gap opening factor $K$ \citep{Kanagawa15} reaches 0.25:
\begin{equation}
    K = q_{\rm p}^2 \hat{h}_{\rm g}^{-5}\alpha^{-1},
\end{equation}
where $q_{\rm p} \equiv m_{\rm p}/M_{\star}$ is the planet-to-star mass ratio. This perturbation was implemented via torque deposition (the angular momentum injection term in Eq. \ref{eq:adv-diff}), following the methodology detailed in Section 2.1.2 of \citet{Lau25}. The torque density exerted by the planet is determined using the empirical gap profile formula from \citet{Duffell20}:
\begin{equation}
    \frac{\Sigma_{\rm g,gap}}{\Sigma_{\rm g,0}} = \left(1 + \dfrac{0.45}{3\pi} \dfrac{\tilde{q}(r)^2}{\alpha \hat{h}_{\rm g}^5} \delta\big(\tilde{q}(r)\big)\right)^{-1},
    \label{eq:gap}
\end{equation}
where $\Sigma_{\rm g,0}$ and $\Sigma_{\rm g,gap}$ denote the gas surface densities before and after gap formation, respectively. For the definitions of the functions $\delta(\tilde{q})$ and $\tilde{q}(r)$, we refer the reader to Eqs. (9) and (18) of \citet{Duffell20}.

\subsubsection{Disk's impact on planets} \label{sssec:met_itact_mig}

The influence of the disk on planetary bodies includes aerodynamic gas drag and gravitational torques, which collectively result in orbital migration and the damping of eccentricity and inclination.

Aerodynamic drag is modeled following \citet{Adachi76}:
\begin{equation}
    \bm{a}_{\text{drag}} = - \left( \frac{3 C_{\text{D}} \rho_{\rm g}(z)}{8 R_{\text{p}} \rho_{\text{s}}} \right) v_{\text{rel}} \bm{v}_{\text{rel}},
\end{equation}
where $C_{\text{D}} = 0.5$ is the drag coefficient \citep{Whipple72}, $\bm{v}_{\text{rel}}$ is the relative velocity between the planetesimal and the gas, and the gas density $\rho_{\rm g}(z)$ at a height $z$ above the midplane is $\rho_{\rm g}(z) = \rho_{\rm g,mid}\exp(-z^2/2H_{\rm g}^2)$.

For migration and damping, we followed the prescriptions of \citet{Ida20}, but adopted the non-isothermal migration timescale from Lee \& Lau (in prep.). This choice avoids numerical instabilities that typically occur near the transition between sub-sonic and super-sonic regimes for specific disk surface density and temperature slopes. The evolution timescales for the semi-major axis $a$, eccentricity $e$, and inclination $i$ are
\begin{align}
    &\tau_a = \frac{t_{\rm wave}}{2 \hat{h}_{\rm g}^2} \left[\frac{C_{\rm P} + 0.5 C_{\rm M}(\hat{e}^2+\hat{i}^2)^{3/2}}{\gamma_{\rm eff}(1+(\hat{e}^2+\hat{i}^2)^2)} -\frac{\Gamma_{\rm c}}{\Gamma_0}\right]^{-1}, \label{eq:tau_a}\\
    &\tau_e = 1.282 t_{\rm wave}  \left[ 1 + \frac{(\hat{e}^2 + \hat{i}^2)^{3/2}}{15} \right],\\
    &\tau_i = 1.838 t_{\rm wave}  \left[ 1 + \frac{(\hat{e}^2 + \hat{i}^2)^{3/2}}{21.5} \right],
\end{align}
where $\hat{e} \equiv e/ \hat{h}_{\rm g}$ and $\hat{i} \equiv i/ \hat{h}_{\rm g}$. The characteristic wave timescale $t_{\rm wave}$ \citep{Tanaka02} is
\begin{equation}
    t_{\rm wave} = q_{\rm p}^{-1}\left(\frac{M_{\odot}}{\Sigma_{\rm g}r^2}\right)\left(\frac{\hat{h}_{\rm g}^4}{\Omega_{\rm K}}\right).
    \label{eq:twave}
\end{equation}
The factors $C_{\rm P}$ and $C_{\rm M}$ are defined as
\begin{align}
    &C_{\rm P} = 2.5 - 0.1p_{\Sigma} +1.7q_T,\\
    &C_{\rm M} = 6(2p_{\Sigma}-q_T+2),
\end{align}
with $p_{\Sigma} = -{\rm d}\ln \Sigma_{\rm g}/{\rm d}\ln r$ and $q_T = -{\rm d}\ln T / {\rm d}\ln r$. The calculations for the effective adiabatic exponent $\gamma_{\rm eff}$ and the normalized corotation torque $\Gamma_{\rm c}/\Gamma_0$ followed \citet{Paardekooper11}. These effects were implemented in the equations of motion in cylindrical coordinates $(r, \phi, z)$:
\begin{equation}
    \bm{a} = - \frac{v_{\mathrm{K}}}{2\tau_a} \bm{e}_{\phi} 
- \frac{v_r}{\tau_e} \bm{e}_r 
- \frac{v_{\phi} - v_{\mathrm{K}}}{\tau_e} \bm{e}_{\phi} 
- \frac{v_z}{\tau_i} \bm{e}_z.
\end{equation}

As noted by \citet{Kanagawa18}, the magnitude of the torque scales linearly with the gas surface density as a planet grows and clears its gap. This leads to a smooth, self-consistent transition from Type I to the high-mass (Type II) migration regime. In our framework, this dependency is naturally included in the calculation of $t_{\rm wave}$ through its $\Sigma_{\rm g}$ term. Combined with our explicit gap opening implementation, this transition is captured automatically, precluding the need for a separate Type II migration treatment.

\subsection{Numerical setup} \label{ssec:met_nume}

\begin{table*}[!htp]
\caption{Summary of model parameters.}
\label{tab:param}
\centering
\begin{tabular}{l l l l l}
\hline\hline
Parameter & Description & Reference & Fiducial value & Tested value\\
\hline
$\alpha$ & Viscosity parameter & Eq. (\ref{eq:alpha}) & $10^{-4}$ & $1.5 \times 10^{-4}$ \\
$v_{\rm frag}$ & Fragmentation velocity & Sec. \ref{sssec:met_disk_dust} & $350\,\rm cm\,s^{-1}$ & $200\,\rm cm\,s^{-1}$ \\
$\delta$ & Small-scale diffusion parameter & Eq. (\ref{eq:Qp}) & $10^{-6}$ & $5\times 10^{-6}$ \\
$r_{\rm infall}$ & Infall location & Eq. (\ref{eq:infall}) & $80\,\rm au$ & $40\,\rm au$ \\
$\kappa$ & Envelope opacity & Eq. (\ref{eq:kappa}) & $0.01\,\rm cm^2\,g^{-1}$ & $0.05\,\rm cm^2\,g^{-1}$ \\
$r_{\rm c}$ & Initial characteristic disk radius & Eq. (\ref{eq:Sig_init}) & $200\,\rm au$ & $50\,\rm au$ \\
$M_{\rm disk}$ & Initial disk mass & Eq. (\ref{eq:Sig_init}) & $0.01\, M_{\odot}$ & $0.02\, M_{\odot}$\tablefootmark{a} \\
\hline
\end{tabular}
\tablefoot{In the fiducial simulations, each parameter is set to its fiducial value. In the parameter tests, all parameters except the one being investigated are held at their fiducial values. \\
\tablefoottext{a}{$\dot{M}_{\rm infall}$ is also doubled when testing $M_{\rm disk}$.}
}
\end{table*}

The dust particle mass grid in \texttt{DustPy} consists of 162 bins, logarithmically spaced from $5.86 \times 10^{-15}$ to $10^8\,{\rm g}$, where the minimum mass corresponds to a particle size of \SI{0.1}{\micro\meter} and the maximum corresponds to 257.4 cm. Our radial grid comprises 201 cells spanning from 3 to $10^3$ au. This includes 153 logarithmically refined cells between the inner boundary of the $N$-body simulation (4 au) and the infall location ($r_{\rm infall} = 80$ au). The $N$-body simulation domain in \texttt{SyMBAp} is defined by trimming five cells from both ends of the \texttt{DustPy} radial grid, resulting in boundaries at 4 au and 793 au. Particles that migrate beyond these limits are removed from the simulation. The $N$-body time step is fixed at $\Delta t \approx 0.54$ yr, which corresponds to $1/15$ of the orbital period at the inner boundary.

To integrate the additional physical effects into \texttt{SyMBAp}, we employed the following operator splitting sequence:
\begin{equation}
    \mathcal{P}^{\Delta t/2}\mathcal{M}^{\Delta t/2}\mathcal{N}^{\Delta t}\mathcal{M}^{\Delta t/2} \mathcal{P}^{\Delta t/2}.
\end{equation}
Here, the operator $\mathcal{P}$ handles pebble accretion, gas accretion, and gap opening; $\mathcal{M}$ accounts for the disk's dynamical impact on the planets; and $\mathcal{N}$ represents the second-order symplectic integrator of \texttt{SyMBAp}. Note that $\mathcal{P}$ and $\mathcal{M}$ operate in heliocentric coordinates, while $\mathcal{N}$ utilizes democratic heliocentric coordinates, necessitating coordinate transformations at each integration step.

The key parameters and their tested values are summarized in Table \ref{tab:param}. We present the results of the fiducial simulations in Section \ref{ssec:res_fidu} and the parameter study in Section \ref{ssec:res_param}. For each configuration, we performed five independent simulations, each covering an evolutionary period of 2 Myr. Our parameter selection was guided by the following physical considerations:

\begin{itemize}
    \item We focused on setups that enable planetesimal formation, guided by the critical conditions identified in \cite{Zhao25}. These conditions require a specific hierarchy of timescales, where the gas diffusion timescale must be significantly longer than both the dust drift and dust growth timescales to allow for efficient local concentration of solids.

    \item Under conditions of low turbulence and high fragmentation velocity, dust growth in the outer disk may be unimpeded by fragmentation. In a pressure bump without such a limit, particles can grow indefinitely into the Stokes II drag regime, which is not supported by \texttt{DustPy}. To ensure numerical and physical consistency, we selected $\alpha$ and $v_{\rm frag}$ such that the fragmentation limit remains applicable across the infall-induced local pressure maximum.

    \item To avoid the Rossby Wave Instability (RWI), which can lead to vortex formation (a high-dimensional process that \texttt{DustPy} cannot capture), we adopted the criterion from \citet{Chang23} based on the relationship between the epicyclic frequency, the radial buoyancy frequency, and $\Omega_{\rm K}$. This ensures the infall-induced pressure bump maintains a moderate radial pressure gradient.

    \item Since disk self-gravity is not modeled in \texttt{DustPy}, we ensured the gas Toomre parameter, $Q = c_{\rm s}\Omega_{\rm K}/\pi G \Sigma_{\rm g}$, remains above 1.5 following the infall to maintain gravitational stability.
\end{itemize}

\subsection{Radiative transfer and synthetic observation} \label{ssec:rad_syn}

To generate observational signatures from our simulation outputs, we utilized the \texttt{RADMC-3D} \citep{radmc} module via the \texttt{DustPyLib} extension \citep{dustpylib}. We performed full radiative transfer calculations, accounting for anisotropic scattering with polarization. While both hemispheres of the $\theta$ grid are required for these calculations, an azimuthal $\varphi$ grid is unnecessary due to the assumed axisymmetry. The radial grid for \texttt{RADMC-3D} is derived from the \texttt{DustPy} grid, with additional local refinement at the disk's inner edge to ensure numerical accuracy. To optimize computational efficiency, the grain size grid is constructed logarithmically between the current minimum and maximum particle sizes, thereby omitting empty mass bins. We adopted the dust opacity model from \citet{Ricci10} and employed $10^7$ photons for the thermal Monte Carlo simulations and $10^6$ photons for the scattering Monte Carlo simulations. The resulting images are generated at a wavelength of $\lambda = 1.3$ mm, corresponding to ALMA Band 6.

The radiative transfer images are subsequently processed with \texttt{SIMIO-continuum} \citep{simio}, which utilizes the \texttt{CASA} \citep{casa} package to produce synthetic ALMA continuum observations. \texttt{SIMIO-continuum} incorporates technical properties identical to existing ALMA datasets, including antenna configuration, integration time, and sky coordinates. The synthetic observations for our fiducial simulations, presented in Section \ref{ssec:syn_fidu}, used the observation of HD 163296 \citep{Isella18} as a template. In Section \ref{ssec:syn_dsharp}, we evaluate the ability of our model to replicate multi-ring configurations similar to those of HD 163296 and AS 209 \citep{Guzman18}, two representative disks from the DSHARP survey \citep{Andrews18}. For these comparisons, the distance and geometry (inclination and position angle) of our simulated disks are matched to the observations. A noise level consistent with the DSHARP data is added.

\section{Simulation results} \label{sec:result}

\subsection{Fiducial setup} \label{ssec:res_fidu}

The stochastic nature of sampling planetesimal masses and orbits, combined with chaotic $N$-body dynamics, results in diverse evolutionary pathways and final planetary system architectures for simulations with identical parameters and initial conditions. We examine the evolution of two representative simulations (Simulation 1 and 2) in detail in Sections \ref{sssec:fidu1} and \ref{sssec:fidu2}, illustrated in Figs. \ref{fig:fidu1} and \ref{fig:fidu2}. The remaining three fiducial simulations are summarized in Section \ref{sssec:fidu_other}. Fig. \ref{fig:history} provides a comparative overview of the evolutionary histories of all fiducial runs.

In this work, we categorize planetesimals into ``generations'' based on their site of origin. The first generation (G1, blue markers in Figs. \ref{fig:fidu1}, \ref{fig:fidu2}, and \ref{fig:history}) originates directly from the initial infall-induced pressure bump. Subsequent generations form within pressure bumps created by the gravitational perturbations of previously formed planets. These generations are ordered chronologically; however, an adjacent numbering does not necessarily imply a direct causal link between every generation.

The early evolution prior to the formation of the first planetesimal involves only \texttt{DustPy} and is identical across all simulations. During the first $5 \times 10^4$ years, late infall enhances the local gas density near $r_{\rm infall}$, creating a pressure maximum that traps both infalling solids and dust drifting inward from the outer disk. Over time, viscous diffusion tends to broaden the pressure bump and drive the pressure maximum slowly inward. By 0.2 Myr, the criteria for planetesimal formation are satisfied at 72 au, triggering the birth of the first planetesimals. The detailed gas and dust dynamics of this stage are further described in \cite{Zhao25}.

\subsubsection{Simulation 1} \label{sssec:fidu1}

\begin{figure}[htp!]
\resizebox{\hsize}{!}{\includegraphics{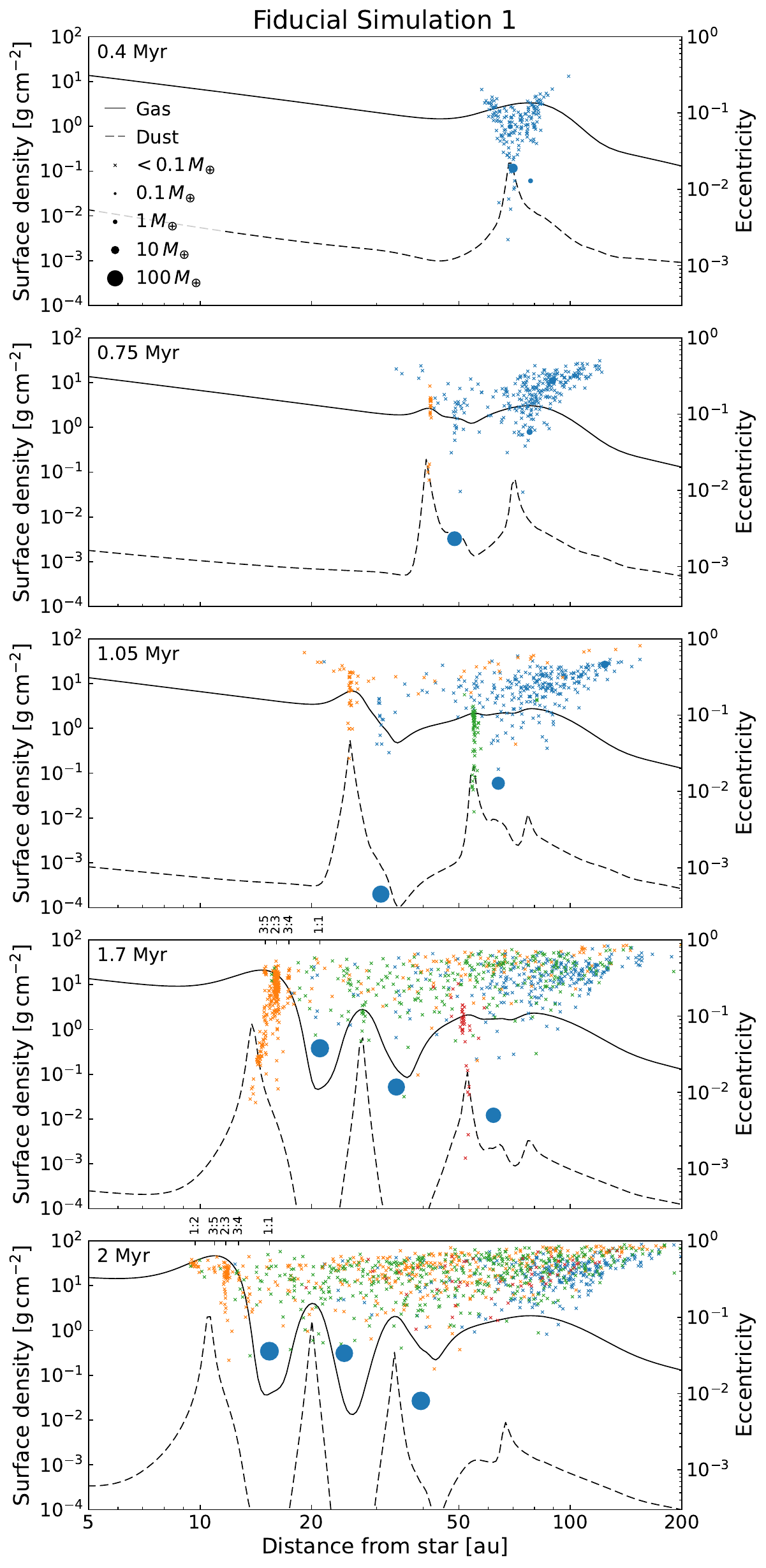}}
\caption{Snapshots of fiducial Simulation 1 at five key epochs. The timestamp for each snapshot is indicated in the upper left corner of each panel. Solid and dashed lines represent the surface densities of gas ($\Sigma_{\rm g}$) and dust ($\Sigma_{\rm d}$), respectively. Scattered markers indicate the semi-major axes ($a$) and eccentricities ($e$) of $N$-body particles. Bodies with $m_{\rm p} < 0.1\, M_{\oplus}$ are marked with ``$\times$,'' while more massive bodies are shown as circles with radii proportional to $m_{\rm p}^{1/3}$. Colors denote different planetesimal generations: blue (G1), orange (G2), green (G3), and red (G4). Ticks on the upper axes denote mean motion resonances relative to the giant planets marked ``1:1.''}
\label{fig:fidu1}
\end{figure}

The top panel of Fig. \ref{fig:fidu1} illustrates Simulation 1 at 0.4 Myr. The infall-induced pressure bump has concentrated dust into a narrow ring, evidenced by a prominent peak in the dust surface density, where G1 planetesimal formation is ongoing. Initial planetesimal masses range from approximately $10^{-9}\, M_{\oplus}$ to $10^{-4}\, M_{\oplus}$. Only three early-forming ($t < 0.25$ Myr), massive ($m_{\rm p,init} > 10^{-5}\, M_{\oplus}$) planetesimals have accreted pebbles efficiently enough to surpass 0.1 $M_{\oplus}$ by this epoch. Less massive planetesimals are scattered by these embryos onto eccentric orbits, where pebble accretion enters the low-efficiency ballistic regime. Collisions in this distant region are rare, and their contribution to planetary growth can be neglected. At this stage, the most massive planet has reached an ice-giant mass (comparable to Uranus), while the other two massive planets remain super-Earths. The infall-induced gas density maximum acts as a migration trap by providing a positive net torque, tethering these planets to the pebble-rich region.

By 0.75 Myr (second panel, Fig. \ref{fig:fidu1}), the primary planet has matured into a gas giant ($76\, M_{\oplus}$) and significantly perturbed the disk. The resulting partial gap divides the original infall-induced pressure bump into two separate planet-induced maxima, effectively splitting the single dust ring. This perturbation breaks the migration trap, allowing the giant to migrate inward. The giant's inward movement steepens the pressure gradient of the inner bump, triggering the formation of the second generation of planetesimals (G2, orange markers). Meanwhile, the two super-Earths in the outer disk have been scattered onto eccentric orbits; though their pebble accretion has slowed, they continue to grow via gas accretion in the cold, infall-replenished environment.

At 1.05 Myr (third panel, Fig. \ref{fig:fidu1}), the giant planet migrates inward and widens the gap as it grows, which sustains the inner pressure bump against viscous dissipation. This allows for persistent G2 formation, in contrast to the rapid decay of infall-induced bumps observed in \cite{Zhao25} when planetary feedback was absent. One of the super-Earths has now grown into a 50 $M_{\oplus}$ gas giant; its eccentricity has been damped as its mass increased ($\tau_e \propto m_{\rm p}^{-1}$). This second giant also carves a partial gap and migrates inward, further concentrating dust in the region between the two giants. Consequently, the third generation of planetesimals (G3, green markers) begins to form in a ``sandwiched'' pressure bump.

By 1.7 Myr (fourth panel, Fig. \ref{fig:fidu1}), a fourth generation (G4, red markers) emerges following the formation of a third giant planet. This giant, which was the other super-Earth in the outer region at 0.75 Myr, underwent the same evolutionary processes as the previous giant: pebble and gas accretion, eccentricity damping, gap opening, and inward migration. Planetesimals formed in these sandwiched regions (G3 and G4) are typically scattered by the adjacent migrating giants shortly after their birth. The inner two giants have now carved deep gaps, significantly slowing their migration. Many newly formed G2 planetesimals are captured into the 3:5, 2:3, and 3:4 mean motion resonances with the innermost giant.

The final configuration at 2 Myr (bottom panel, Fig. \ref{fig:fidu1}) reveals a system of three giant planets (163, 132, and 155 $M_{\oplus}$) orbiting at 15.4, 24.5, and 39.5 au, near a 1:2:4 resonance chain. Three distinct dust rings are present: one interior to the innermost planet and two located between the planetary orbits. The synthetic observations of this architecture are discussed in Section \ref{sssec:syn_fidu1}. G1 planetesimals that remained small are primarily scattered into the outer disk on highly eccentric orbits. In contrast, the G2 population is distributed broadly due to their prolonged formation period and the long-range migration of the planets. The number of resonant G2 bodies has decreased since 1.7 Myr, likely due to a dynamical instability triggered as the outermost giant approached the 1:2:4 commensurability. G3 and G4 planetesimals are uniformly scattered to high eccentricities. In the outer disk, all four generations are now well-mixed, representing an age span of 1.8 Myr. None of the later generations successfully grew into massive bodies.

\subsubsection{Simulation 2} \label{sssec:fidu2}

\begin{figure}[htp!]
\resizebox{\hsize}{!}{\includegraphics{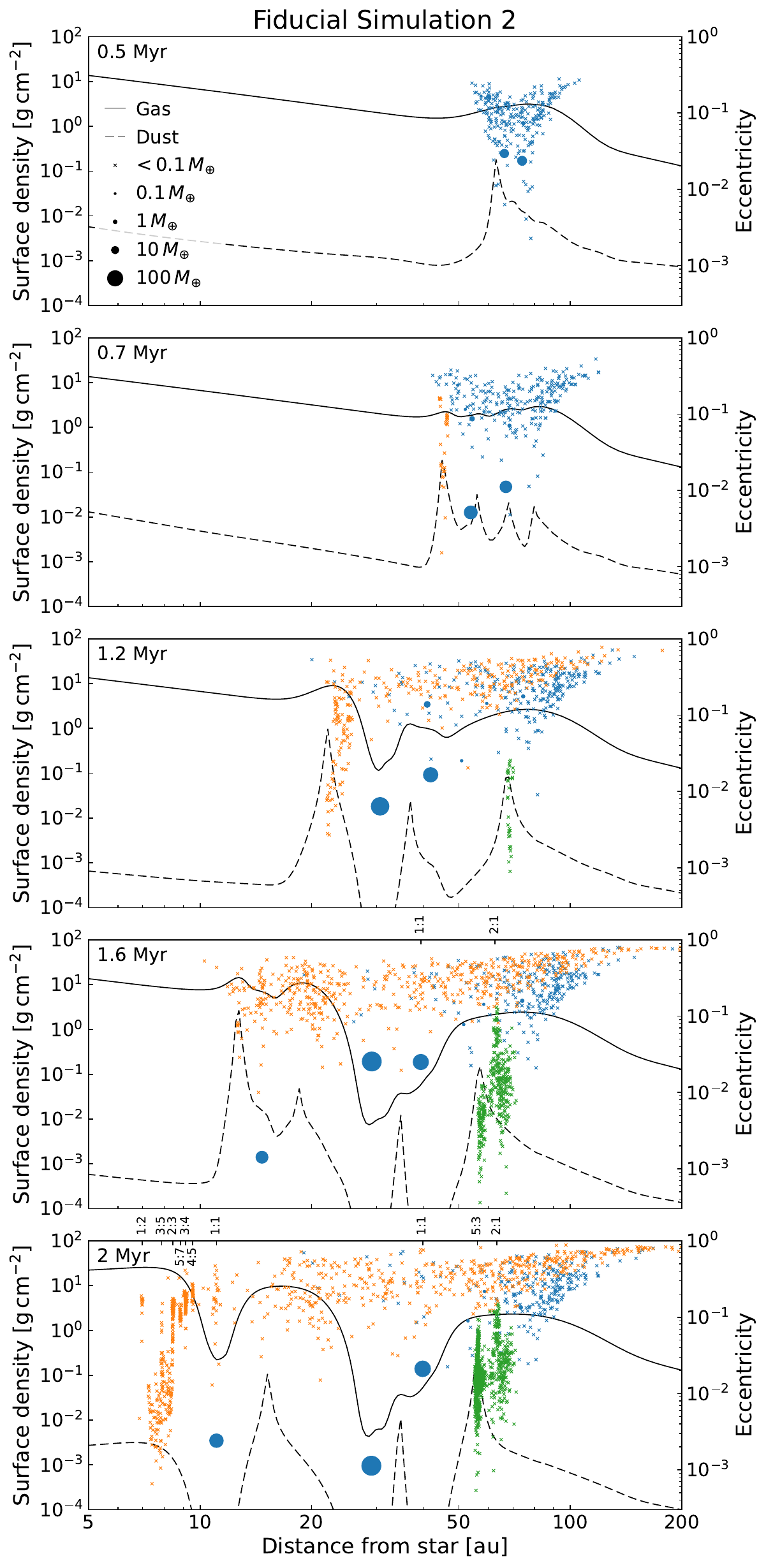}}
\caption{Snapshots of fiducial Simulation 2 at five key epochs. Notation and symbology are identical to Fig. \ref{fig:fidu1}.}
\label{fig:fidu2}
\end{figure}

The formation mechanisms for G1 and G2 planetesimals in Simulation 2 are qualitatively similar to those in Simulation 1. By 0.5 Myr (top panel, Fig. \ref{fig:fidu2}), two protoplanets have reached ice-giant masses, one has grown into a super-Earth, and three others have surpassed 0.1 $M_{\oplus}$. By 0.7 Myr (second panel, Fig. \ref{fig:fidu2}), the feedback from these giants has triggered G2 formation at the inner pressure bump.

A divergence in evolutionary pathways becomes evident by 1.2 Myr (third panel, Fig. \ref{fig:fidu2}). Unlike the sandwiched G3 formation seen in Simulation 1, the G3 planetesimals here form in the pressure bump exterior to the outermost giant, sustained solely by the inward-drifting dust from the outer disk. Although a pressure bump exists between the two giants, it fails to trigger planetesimal formation due to a localized deficiency in dust. Meanwhile, a super-Earth currently co-orbital with the outer giant is frequently scattered, yet it continues to accrete gas steadily despite its unstable orbital configuration.

By 1.6 Myr (fourth panel, Fig. \ref{fig:fidu2}), this super-Earth has been scattered inward. Upon entering the inner pressure bump, its eccentricity is rapidly damped by several orders of magnitude due to the high local gas density. This stabilization, combined with the abundant supply of solids and gas, facilitates rapid growth; the planet reaches $47\, M_{\oplus}$ and opens a partial gap, shifting the G2 formation site further inward. The two original giants now share a broad, deep gap that is impermeable to dust. In the outer disk, a significant fraction of G3 planetesimals are captured into the 2:1 mean motion resonance with the outermost giant. Notably, the majority of G3 bodies maintain low eccentricities because no giant planet has migrated through their formation region to scatter them.

The final configuration at 2 Myr (bottom panel, Fig. \ref{fig:fidu2}) consists of three giants (68, 196, and 107 $M_{\oplus}$) orbiting at 11.1, 29.0, and 39.9 au. The innermost pressure bump has dissipated after losing the dynamical support of the innermost giant, which has effectively stalled migration following the opening of a deep gap. A portion of the most recently formed G2 planetesimals is captured in various resonances with the innermost giant, while the uncaptured ones further afield remain at low eccentricities. The G3 population also exhibits generally low eccentricities, with many members trapped in 5:3 and 2:1 resonances with the outermost giant. The preservation of these low-eccentricity populations is a key distinction from Simulation 1. However, their growth remains limited by the depleted pebble flux at this late stage. The synthetic observations of this simulation are presented in Section \ref{sssec:syn_fidu2}.

Based on Simulations 1 and 2, we categorize planet-induced, multigenerational planetesimal formation into three distinct scenarios:
\begin{enumerate}
\item Inner-side formation: Planetesimals form interior to the innermost giant. Those born while the giant is migrating rapidly are scattered across the disk, whereas those born after migration slows are either captured into resonance or remain at low eccentricities.
\item Sandwiched formation: Planetesimals form between two giants. These are invariably scattered onto high-eccentricity orbits by the adjacent migrating giants.
\item Outer-side formation: Planetesimals form exterior to the outermost giant. These populations typically remain at low eccentricities and are susceptible to resonance capture.
\end{enumerate}

The top-left and top-middle panels of Fig. \ref{fig:history} provide an overview of the evolutionary history of Simulations 1 and 2, respectively. Each ``$\times$'' symbol marks the specific formation time and radial location of an individual planetesimal; these are clustered along distinct tracks that define the different generations. Apart from the initial infall-induced pressure bump, the emergence of each subsequent formation site is associated with the feedback of a growing giant planet (indicated by gray lines). Notably, the model shows that planetesimal formation can occur simultaneously at multiple, widely separated radii.

\subsubsection{Other fiducial simulations} \label{sssec:fidu_other}

\begin{figure*}[htp!]
\resizebox{\hsize}{!}{\includegraphics{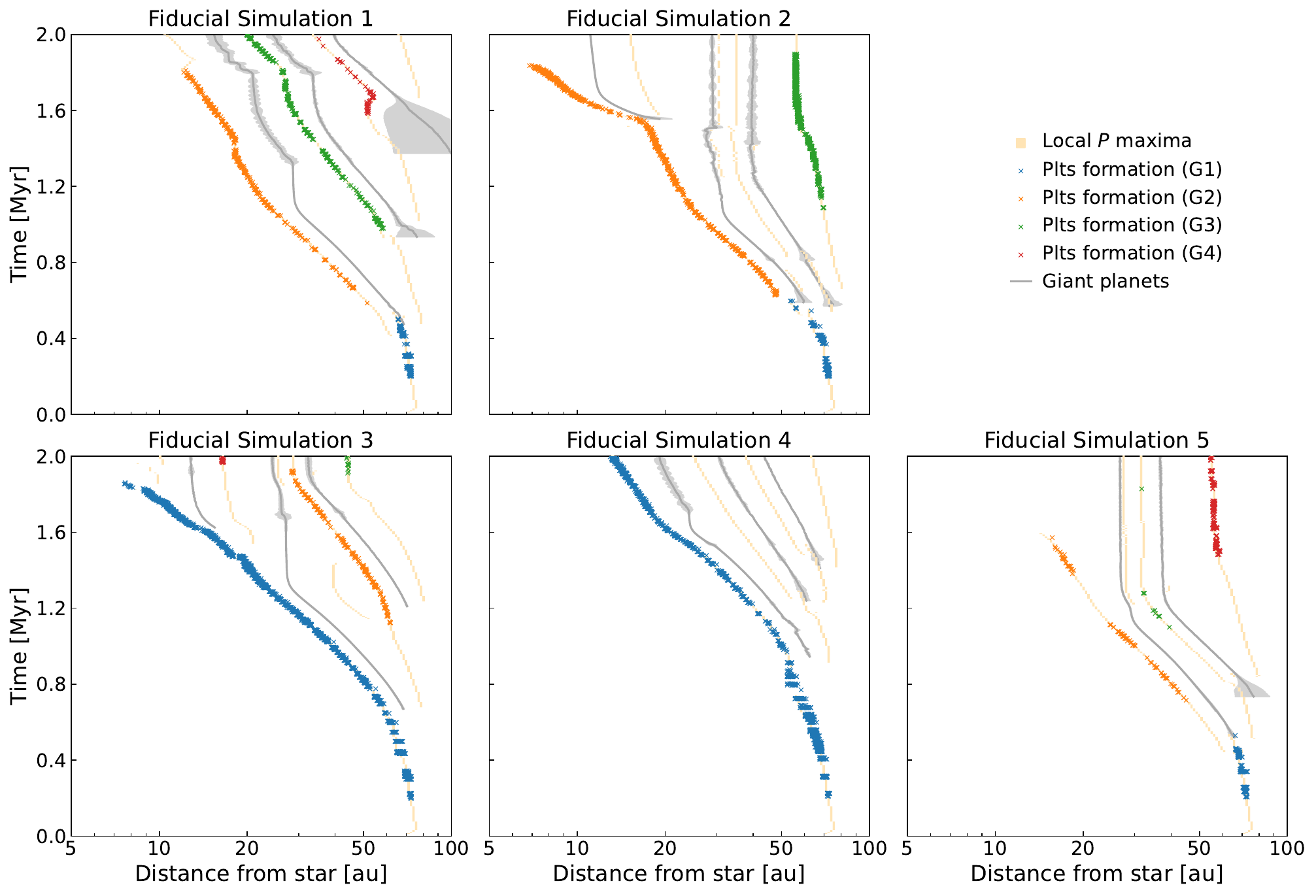}}
\caption{Evolutionary history of local pressure maxima, discrete formation events, and giant planet orbits across all fiducial simulations. Radial cells containing local pressure maxima are indicated in light orange. Individual planetesimal formation events are marked with ``$\times$,'' with colors distinguishing generations as in Figs. \ref{fig:fidu1} and \ref{fig:fidu2}. Gray lines and shaded regions visualize the semi-major axes and eccentricities of giant planets with $m_{\rm p} > 30\, M_{\oplus}$.}
\label{fig:history}
\end{figure*}

While the detailed snapshots for Simulations 3, 4, and 5 are not illustrated, their evolutionary overviews are provided in the bottom row of Fig. \ref{fig:history}. The distribution of semi-major axes and masses for all massive bodies ($m_{\rm p} > 0.1\, M_{\oplus}$) at the end of the 2-Myr fiducial simulations is displayed in the top-left panel of Fig. \ref{fig:param}.

In Simulation 3, the primary giant planet reaches the gap opening threshold significantly later than in Simulations 1 and 2. By the time this planet begins to perturb the gas, the infall-induced pressure maximum has already diffused considerably inward. Consequently, the formation of G1 planetesimals remains uninterrupted, following an inward migration pattern similar to the G2 populations in previous runs. By the end of the simulation, three giant planets ($> 30\, M_{\oplus}$) are present. The innermost giant originates from the G1 population but does not emerge until 1.4 Myr. Subsequent generations follow various pathways: G2 forms in a sandwiched region between the outer two giants, G3 forms on the outer edge of the outermost giant's gap, and G4 emerges in a sandwiched region between the inner pair of giants.

Simulation 4 exhibits even slower initial giant planet growth. Planet-induced pressure bumps are not established until 0.9 Myr, approximately 0.2 Myr after the original infall-induced bump has dissipated via diffusion. This temporal gap in pressure support proves critical; during the period when the disk lacks a trapping mechanism, a significant volume of dust drifts into the inner disk, although planetesimal formation in this dust ring is not interrupted. This leads to a localized deficiency in the dust budget, which ultimately suppresses planetesimal formation in the newly created outer pressure bumps.

Simulation 5 serves as a comprehensive example containing all three formation scenarios identified in Section \ref{sssec:fidu2}. G1 planetesimals emerge from the initial infall-induced bump, G2 forms interior to the innermost giant, G3 represents a sandwiched population between existing giants, and G4 forms exterior to the outermost giant.

The top-left panel of Fig. \ref{fig:param} reveals a clear architectural trend: planets in the 0.1--10 $M_{\oplus}$ range are generally located at larger radii than their more massive counterparts. These medium-mass bodies are initially tethered by the infall-induced migration trap and are subsequently prevented from migrating inward by the dynamical barriers (gaps and resonances) created by the interior giant planets. In disks subject to late infall as parameterized in Table \ref{tab:param}, the formation of giant planets exceeding 100 $M_{\oplus}$ is a robust outcome. Multigenerational formation is consistently enabled by planet-disk interactions, provided that the initial giant planet grows rapidly enough to maintain pressure support before the infall-induced bump dissipates.

\subsection{Effects of parameters} \label{ssec:res_param}

\begin{figure*}[htp!]
\resizebox{\hsize}{!}{\includegraphics{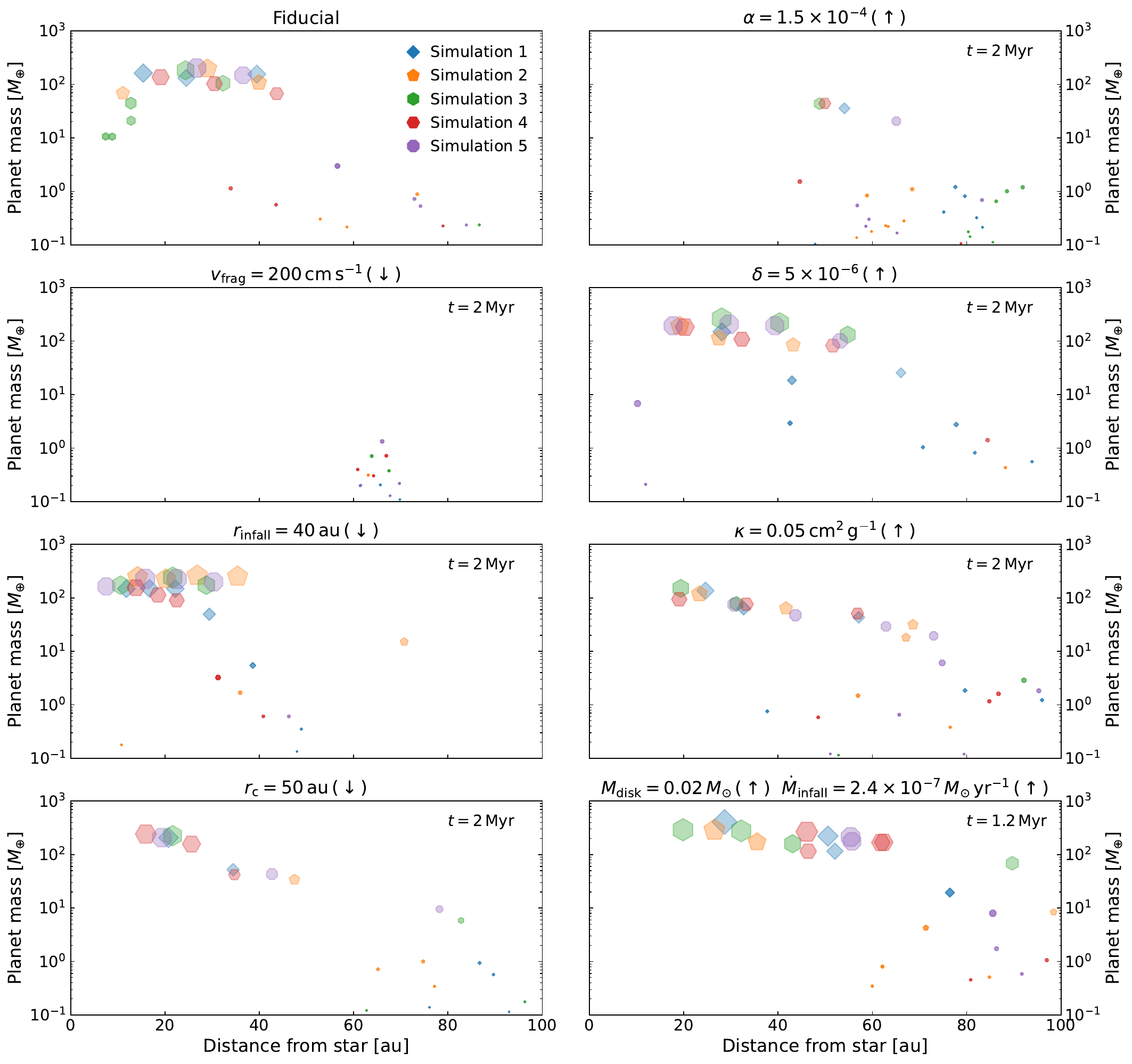}}
\caption{Semi-major axes and masses of planets above $0.1\, M_{\oplus}$ for various parameter setups. The timestamps for the displayed results are located in the top-right corner of each panel. Parameters modified relative to their fiducial values are indicated at the top of each panel, and the arrows in parentheses indicate that the parameter value is being increased or decreased. Different markers distinguish individual simulations within the same setup. Marker sizes are proportional to $m_{\rm p}^{1/3}$, and transparency indicates the mass fraction of the gaseous envelope, with more transparent markers representing gas-rich giants.}
\label{fig:param}
\end{figure*}

We investigate the sensitivity of our model to several key parameters by varying each individually while holding others at their fiducial values. The resulting semi-major axis and mass distributions of massive bodies ($m_{\rm p} > 0.1\, M_{\oplus}$) are summarized in Fig. \ref{fig:param}.

When the turbulent viscosity parameter $\alpha$ is increased to $1.5 \times 10^{-4}$, the enhanced turbulence accelerates the dissipation of the infall-induced pressure bump. Additionally, it reduces the pebble accretion efficiency $\varepsilon_{\rm PA}$ by increasing the pebble scale height and lowering the midplane settling probability. Under these conditions, each simulation produces at most one giant planet ($> 10\, M_{\oplus}$), and none reach $50\, M_{\oplus}$ within 2 Myr. Consequently, multigenerational planetesimal formation is suppressed due to the lack of significant planetary feedback on the disk.

Reducing the fragmentation velocity $v_{\rm frag}$ to $200\,{\rm cm\,s^{-1}}$ limits the maximum grain size through particle fragility. Smaller dust grains are less susceptible to pressure-trap concentration, thereby reducing the mass available for both planetesimal formation and pebble accretion. In this regime, planets fail to grow beyond the super-Earth stage.

Increasing the small-scale dust diffusion parameter $\delta$ to $5 \times 10^{-6}$ increases the characteristic scale of the gravitationally unstable dust clumps formed by the streaming instability. While the total mass available for planetesimal formation remains constant, the resulting population consists of fewer but initially more massive planetesimals. Compared to fiducial runs, a similar number of gas giants form, typically with higher final masses. Furthermore, some later-generation planetesimals possess enough initial mass to undergo efficient pebble accretion, such as the $19\,M_{\oplus}$ planet at 43 au in Simulation 1 and the $7\,M_{\oplus}$ planet at 10 au in Simulation 5.

Shifting the infall location $r_{\rm infall}$ inward to 40 au increases the local gas and dust surface densities at the formation sites. This facilitates more rapid planetary growth but also shortens migration timescales. The resulting planetary systems are more compact and feature more massive giants. The reduced orbital spacing frequently leads to multiple giant planets sharing a single, broad gap.

Increasing the envelope opacity $\kappa$ to $0.05\,{\rm cm^2\,g^{-1}}$ slows the cooling and contraction of the gaseous envelope. This delays the onset of runaway gas accretion, gap opening, and inward migration. In several cases, the infall-induced pressure bump dissipates before planetary feedback can establish new traps, resulting in significant dust loss to the inner disk. While multigenerational planetesimal formation still occurs, its productivity is markedly reduced, and the resulting giant planets have lower masses and reside at larger orbital radii.

Decreasing the initial characteristic disk radius $r_{\rm c}$ to 50 au places the infall location in the outer disk beyond $r_{\rm c}$. Because the majority of the initial dust mass is concentrated within $r_{\rm c}$, it cannot be trapped by the infall-induced bump. Furthermore, the steep initial pressure gradient at the infall radius prevents the maintenance of a robust positive pressure gradient after infall. As a result, gas giant frequency drops and multigenerational planetesimal formation is entirely inhibited.

Finally, we performed an experiment where both the initial disk mass $M_{\rm disk}$ and the infall rate $\dot{M}_{\rm infall}$ were doubled. This setup maintains the pressure gradient parameter $\eta$, preserving dust-trapping efficiency while increasing the total solid budget. The enhanced dust content accelerates pebble accretion, while the higher gas density promotes rapid migration and eccentricity damping. The bottom-right panel of Fig. \ref{fig:param} displays the results at 1.2 Myr, where several planets have grown into Jupiter-mass giants. Additionally, the efficient damping allows for multiple pairs of planets to stay in co-orbital resonances. However, the combination of high masses and rapid migration eventually triggers dynamical instabilities in four out of five simulations after 1.2 Myr. Since our gap opening and accretion prescriptions assume a quasi-steady disk, the model results become unreliable once these chaotic orbital crossings begin.

We also tested a reduced infall rate ($\dot{M}_{\rm infall} = 6 \times 10^{-8}\,M_{\odot}\,{\rm yr}^{-1}$) and a lower infall dust-to-gas ratio ($Z_{\rm infall} = 10^{-4}$). Because both cases simply result in suppressed planetary growth---an effect analogous to that produced by a reduction in fragmentation velocity---their detailed figures and discussions are omitted for brevity.

\section{Synthetic observations} \label{sec:synobs}

\subsection{Representative fiducial simulations} \label{ssec:syn_fidu}

\begin{figure*}[!htp]
\resizebox{\hsize}{!}{\includegraphics{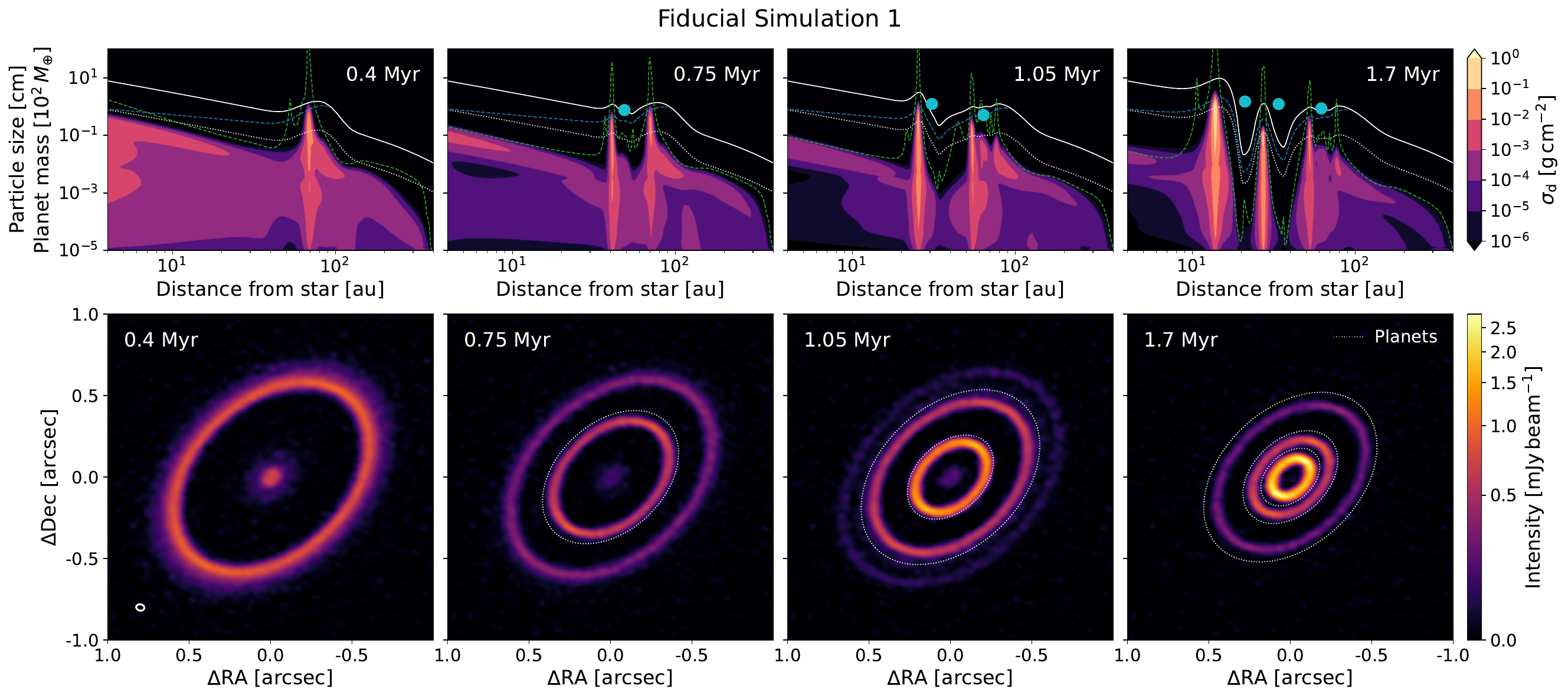}}
\caption{Dust distribution and synthetic observations of Fiducial Simulation 1. Top panels: Grid-independent dust surface density ($\sigma_{\rm d}$) vs. radial distance and particle size. Blue circles indicate planets with $m_{\rm p} > 30\,M_{\oplus}$. White solid/dotted lines denote ${\rm St} = 1$ and $0.1$, while blue and green lines show the fragmentation and drift limits, respectively. Bottom panels: Synthetic 1.3-mm continuum images. The solid ellipse (bottom-left) indicates the beam size. Dotted ellipses represent the orbits of giant planets ($m_{\rm p}> 30\,M_{\oplus}$) based on their semi-major axes.}
\label{fig:fidu_obs1}
\end{figure*}

We generated synthetic observations for Fiducial Simulations 1 and 2 at the same epochs discussed in Sections \ref{sssec:fidu1} and \ref{sssec:fidu2}. The top rows of Figs. \ref{fig:fidu_obs1} and \ref{fig:fidu_obs2} illustrate the dust surface density ($\sigma_{\rm d}$) as a function of radial distance and particle size. Contours of constant Stokes number (white lines) reflect the gas surface density profile, as particle size scales with $\Sigma_{\rm g}$ for a fixed St. The bottom rows of these figures present synthetic 1.3-mm (ALMA Band 6) continuum images. For these visualizations, the disks are assumed to be located at the distance of HD 163296 (101 pc) with an identical orientation ($i=46^{\circ}$, ${\rm PA}=133^{\circ}$), and observed using the technical configuration of the DSHARP survey \citep{Isella18}.

\subsubsection{Synthetic observations: Simulation 1} \label{sssec:syn_fidu1}

At 0.4 Myr, the dust concentrated within the infall-induced pressure bump has produced a bright ring in the 1.3-mm continuum, spanning approximately 60 to 90 au. Despite active planetesimal formation and pebble accretion, the ring retains $15.3\,M_{\oplus}$ of dust within this visible range. The outer edge of the ring appears extended due to the continuous influx of drifting dust from the outer disk. The azimuthally averaged intensity peaks at 69 au with $I_{\rm max,avg} = 0.83\,{\rm mJy\, beam^{-1}}$ ($0.38\,{\rm Jy\, arcsec^{-2}}$). Given the midplane temperature $T(69\,{\rm au}) = 26.6\,{\rm K}$ assumed in our model, the maximum optical depth is
\begin{equation}
    \tau_{\rm max} = -\ln \left(1 - \frac{I_{\rm max,avg}}{B_{\nu}(T)} \right) = 0.61.
\end{equation}
This value is slightly higher than the typical optical depths ($0.2$--$0.5$) reported for DSHARP rings \citep{Dullemond18}. The marginally optically thick rings observed in the DSHARP survey can potentially be explained by the streaming instability regulating the midplane dust-to-gas ratio to unity \citep{Stammler19}. While our model supports this mechanism, this specific ring achieves a higher optical depth due to the high dust supply from infall and our conservative planetesimal formation efficiency ($\zeta = 10^{-6}$), allowing the midplane dust-to-gas ratio to exceed unity.

By 0.75 Myr, the giant planet has bisected the original dust ring into two distinct visible rings. Because the inner pressure bump's survival depends on the giant's gap opening and inward migration, the planet's orbit lies closer to the inner ring than the outer one. This finding challenges the common assumption that a gap opening planet should be centered between two symmetric rings. The inner ring ($37$--$47$ au) contains $3.0\,M_{\oplus}$ of dust with $\tau_{\rm max} = 0.29$, while the outer ring ($66$--$82$ au) contains $5.6\,M_{\oplus}$ with $\tau_{\rm max} = 0.20$. Active planetesimal formation is restricted to the inner ring, explaining its higher optical depth.

By 1.05 Myr, the second giant planet has matured and migrated inward, sharpening the previous outer ring. A third, optically thin ring ($\tau_{\rm max} = 0.07$) has emerged exterior to this second planet, containing $1.4\,M_{\oplus}$ of dust ($73$--$83$ au) due to the pile-up of drifting material. Weak emission is detected near the planet's orbit, as the gap is not yet deep enough to fully clear the dust. The innermost ring has contracted following the first planet's migration. Despite the consumption of solids by planetesimal formation, the ring's mass remains stable ($3.1\,M_{\oplus}$ within $21$--$31$ au) because the migrating pressure bump effectively ``sweeps up'' dust from the inner disk. This contraction increases the dust-to-gas ratio, raising $\tau_{\rm max}$ to 0.51. Meanwhile, planetesimal formation has commenced in the middle ring, where $\tau_{\rm max}$ has increased to 0.33.

At 1.7 Myr, the outermost ring has been compressed by the third giant and has accumulated $3.0\,M_{\oplus}$ ($48$--$58$ au). As it becomes marginally optically thick ($\tau_{\rm max} = 0.14$), planetesimal formation begins here as well. Although the third giant creates an additional trap on its outer edge, the dust in the outer disk has been largely depleted, preventing the formation of further visible rings. The innermost ring continues to compact, reaching a peak intensity of $2.04\,{\rm mJy\, beam^{-1}}$ ($0.93\,{\rm Jy\, arcsec^{-2}}$) and an optical depth of 0.57.

We omit the synthetic image at 2 Myr, as the primary difference from the 1.7 Myr epoch is a further contraction of the ring radii caused by ongoing planet migration.

\subsubsection{Synthetic observations: Simulation 2} \label{sssec:syn_fidu2}

\begin{figure*}[!htp]
\resizebox{\hsize}{!}{\includegraphics{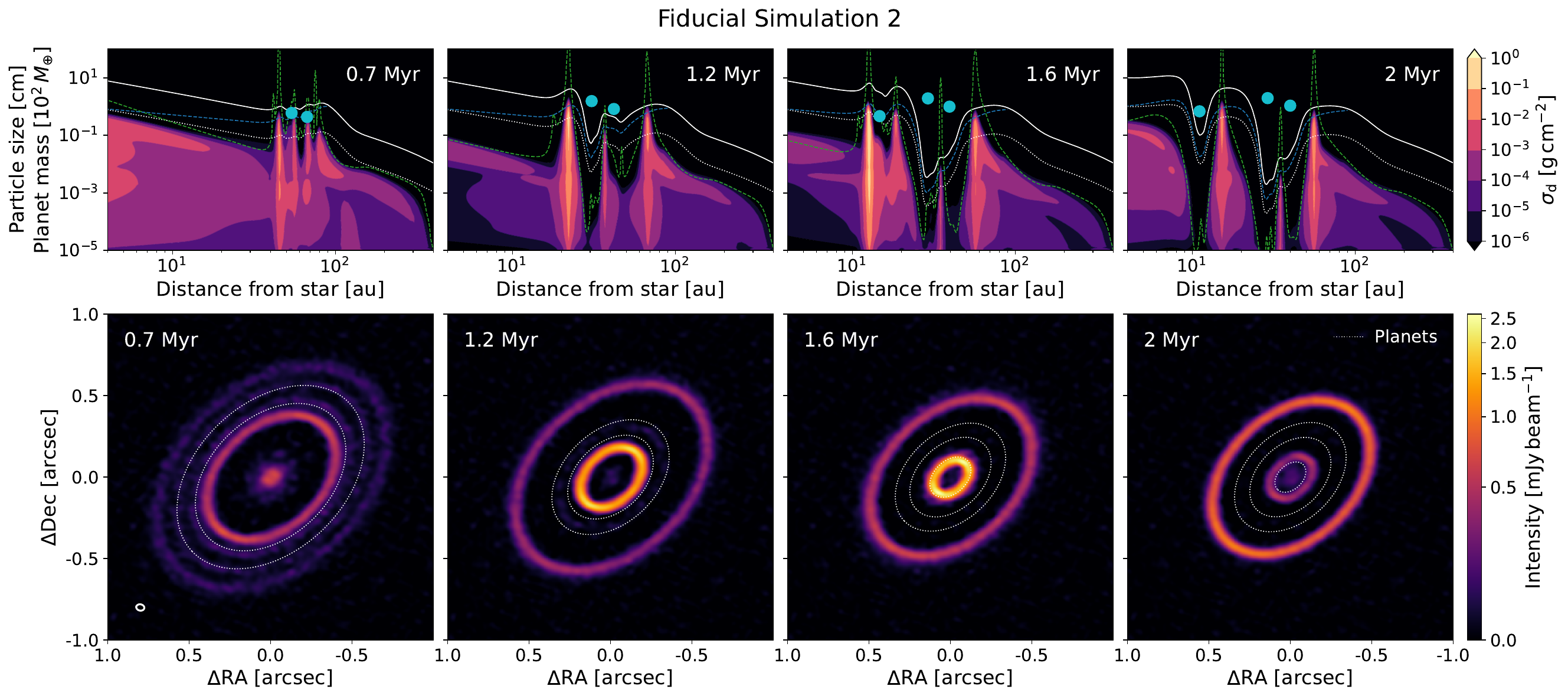}}
\caption{Dust distribution (top panels) and synthetic observations (bottom panels) of Fiducial Simulation 2 at four key epochs. Notations and contour definitions are identical to those in Fig. \ref{fig:fidu_obs1}.}
\label{fig:fidu_obs2}
\end{figure*}

The early gas and dust evolution in Simulation 2 is consistent with Simulation 1, characterized by a local pressure maximum that migrates slowly inward via gas viscous evolution. The infall-induced dust ring at 0.5 Myr is morphologically similar to the 0.4 Myr snapshot of Simulation 1, although it has a slightly more inward-shifted radius of maximum intensity; it is thus omitted from Fig. \ref{fig:fidu_obs2}.

By 0.7 Myr, the simultaneous emergence of two giant planets induces significant fluctuations in the gas surface density. This transforms the single infall-induced dust ring into a bright inner ring ($\tau_{\rm max} = 0.26$ at 45 au), where planetesimal formation is ongoing, and three faint outer rings ($\tau_{\rm max} < 0.06$). The two intermediate rings currently coincide with the planets' orbits; these features vanish rapidly as the giants deepen their gaps. Meanwhile, the outermost ring continues to accumulate drifting material from the outer disk.

At 1.2 Myr, the outer ring has collected sufficient dust to reach an optical depth of 0.22, triggering the onset of planetesimal formation. The giant planets have carved a wide, deep gap between the two primary bright rings. Within this gap, only barely detectable emission is present ($\tau_{\rm max} = 0.02$), corresponding to a mere $0.3\,M_{\oplus}$ of residual dust.

By 1.6 Myr, the emergence of a third giant planet in the inner disk has driven the innermost ring further inward. Because the third planet’s mass is not yet sufficient to fully clear its vicinity, its orbit overlaps with the inner ring. The two original giants have nearly stalled their migration. In the outer disk, the ring radius has contracted as gas diffusion shifts the local pressure maximum inward until it is halted by the outer edge of the planetary gap. The brightness of this outer feature is enhanced ($\tau_{\rm max} = 0.29$) due to continued dust pile-up.

At 2 Myr, the pressure bump interior to the innermost giant has dissipated due to viscous diffusion, releasing its dust toward the central star and leaving only faint emission. A new ring has begun to collect dust ($0.3\,M_{\oplus}$) just exterior to the innermost giant's orbit, though it remains optically thin ($\tau_{\rm max} = 0.10$) and has insufficient dust for planetesimal formation. Conversely, the outermost ring has become significantly more prominent ($\tau_{\rm max} = 0.52$). The deep planetary gaps now block the inward dust flux so effectively that $8\,M_{\oplus}$ of solids are trapped in this single feature.

In summary, the growth and migration of giant planets, which are themselves products of the initial infall-induced ring, drive the transformation of a single ring into a multi-ring system. In our model, planetesimal-forming rings generally exhibit optical depths $\tau_{\rm max} \gtrsim 0.2$. This result is in good agreement with both the measured values from the DSHARP survey \citep{Dullemond18} and the theoretical predictions of regulation by the streaming instability \citep{Stammler19}.

\subsection{Reproduction of DSHARP rings} \label{ssec:syn_dsharp}

Multi-ring disks are ubiquitous in high-resolution ALMA continuum surveys \citep[e.g.,][]{Andrews18,Long18}, yet the origin of these configurations remains a subject of intense debate. Given our model’s ability to generate multi-ring architectures via planetary feedback, we attempted to reproduce the outer rings of two representative disks from the DSHARP survey: HD 163296 \citep{Isella18} and AS 209 \citep{Guzman18}. Our simulations began with an infall-induced dust trap that triggers the formation of giant planets. We adopted the stellar properties (mass, luminosity, effective temperature) from \citet{Oberg21}, setting the initial disk mass to 1\% of the stellar mass and the infall rate to $2.4 \times 10^{-7}\,M_{\odot}\,{\rm yr^{-1}}$. Infall locations were chosen based on the observed ring radii, and fragmentation velocities were adjusted to ensure that the fragmentation limit encompasses the infall-induced trap (as discussed in Section \ref{ssec:met_nume}). All other parameters remain identical to the fiducial setup.

\subsubsection{HD 163296}

\begin{figure}[!htp]
\resizebox{\hsize}{!}{\includegraphics{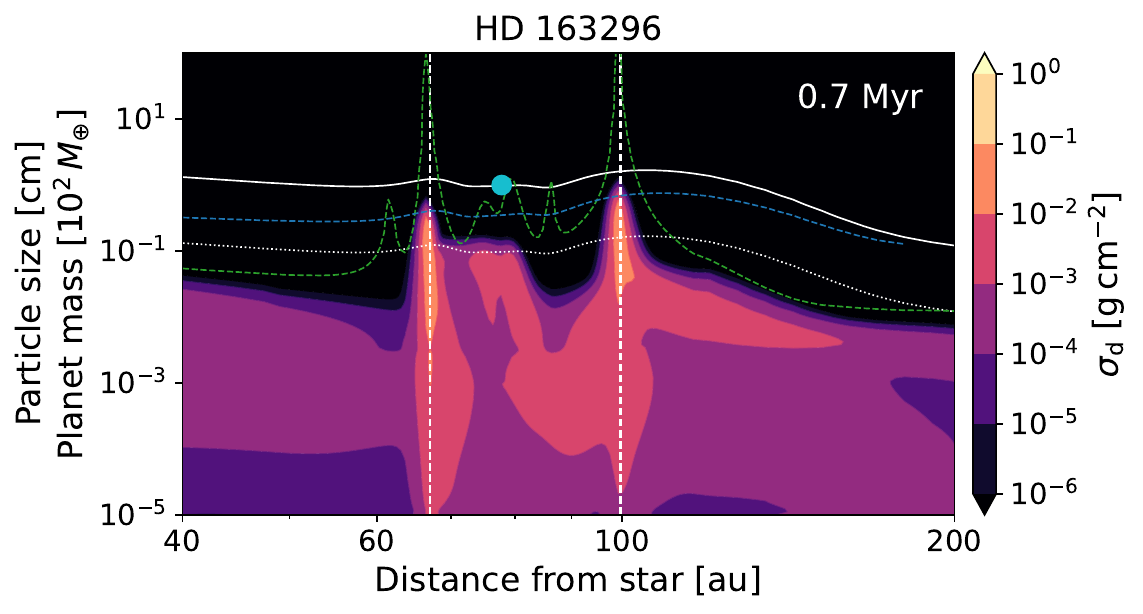}}
\resizebox{\hsize}{!}{\includegraphics{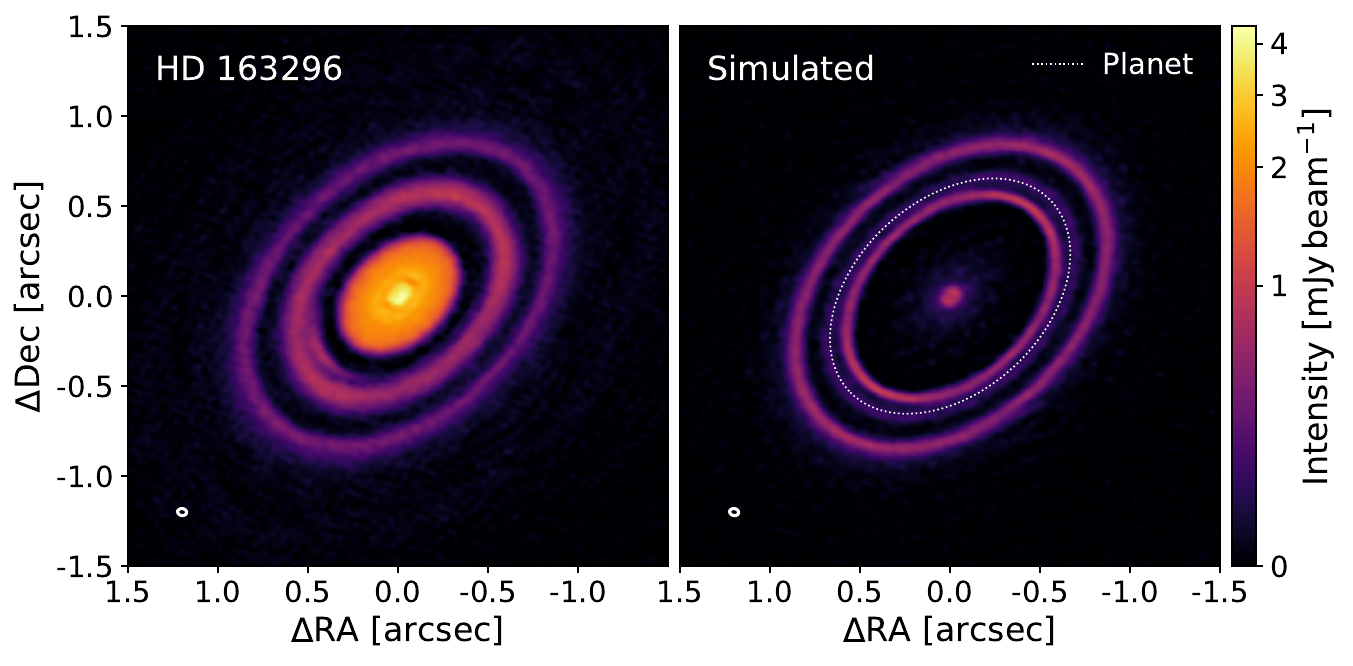}}
\caption{Dust distribution (top) and synthetic 1.3-mm continuum observation (bottom right) at 0.7 Myr, compared to the DSHARP observation of HD 163296 (bottom left; \citealp{Andrews18,Isella18}). Vertical lines in the top panel indicate the radial locations of the rings of interest. Other notation follows Fig. \ref{fig:fidu_obs1}.}
\label{fig:HD163296}
\end{figure}

The disk of HD 163296 features two prominent rings at 67 and 100 au. The central star is a 2 $M_{\odot}$ Herbig Ae star ($T_{\rm eff} = 9332$ K, $R_{\star} = 1.6\,R_{\odot}$). We set the infall location to $r_{\rm infall} = 110\,{\rm au}$.

Fig. \ref{fig:HD163296} (top) shows the dust distribution at 0.7 Myr after the onset of late infall. A $99\,M_{\oplus}$ giant planet at 78 au has opened a partial gap in the infall-induced pressure bump, redistributing the trapped dust into two secondary maxima on either side of its orbit. These traps migrate inward with the planet and align with the observed B67 and B100 rings of HD 163296. In the inner ring, a second generation of planetesimals has already begun to form.

The synthetic 1.3-mm continuum image (Fig. \ref{fig:HD163296}, bottom) clearly reproduces both rings. The simulated B100 ring (90--110 au) contains $12.3\,M_{\oplus}$ of dust and reaches a peak intensity of $0.28\,{\rm Jy\, arcsec^{-2}}$, corresponding to $\tau_{\rm max} = 0.32$. This is marginally brighter than the DSHARP observation ($0.21\,{\rm Jy\, arcsec^{-2}}$; $\tau_{\rm max} = 0.23$ measured with our temperature model). Interestingly, \citet{Dullemond18} estimated $\tau_{\rm max} = 0.33$ for this ring using a different temperature model, which matches our result nearly perfectly. While \cite{Dullemond18} estimated a much higher dust mass ($43.6\,M_{\oplus}$) using the \citet{Birnstiel18} opacity model, our use of the \citet{Ricci10} model achieves comparable brightness with significantly less dust. The maximum grain size in our model (0.68 cm) remains well below the stability limit (9.5 cm) provided by \cite{Dullemond18}.

The simulated B67 ring (60--75 au) contains $5.7\,M_{\oplus}$ of dust with an intensity of $0.32\,{\rm Jy\, arcsec^{-2}}$ ($\tau_{\rm max} = 0.29$), slightly dimmer than the observed values ($0.36\,{\rm Jy\, arcsec^{-2}}$; $\tau_{\rm max} = 0.32$). Both simulated rings appear narrower than in the observations, a discrepancy likely caused by our assumption of low dust diffusivity or neglect of planets' hydrodynamical effects on dust \citep{Bi21,Binkert21}. Because the giant planet has not yet cleared its vicinity, residual emission from $2.8\,M_{\oplus}$ of dust between 75 and 90 au creates a ``shoulder'' at the outer edge of the B67 ring not seen in the observational data. Furthermore, while the real HD 163296 possesses a bright inner disk, our simulated inner disk is depleted as we did not include inner-disk pressure bumps.

\subsubsection{AS 209}

\begin{figure}[!htp]
\resizebox{\hsize}{!}{\includegraphics{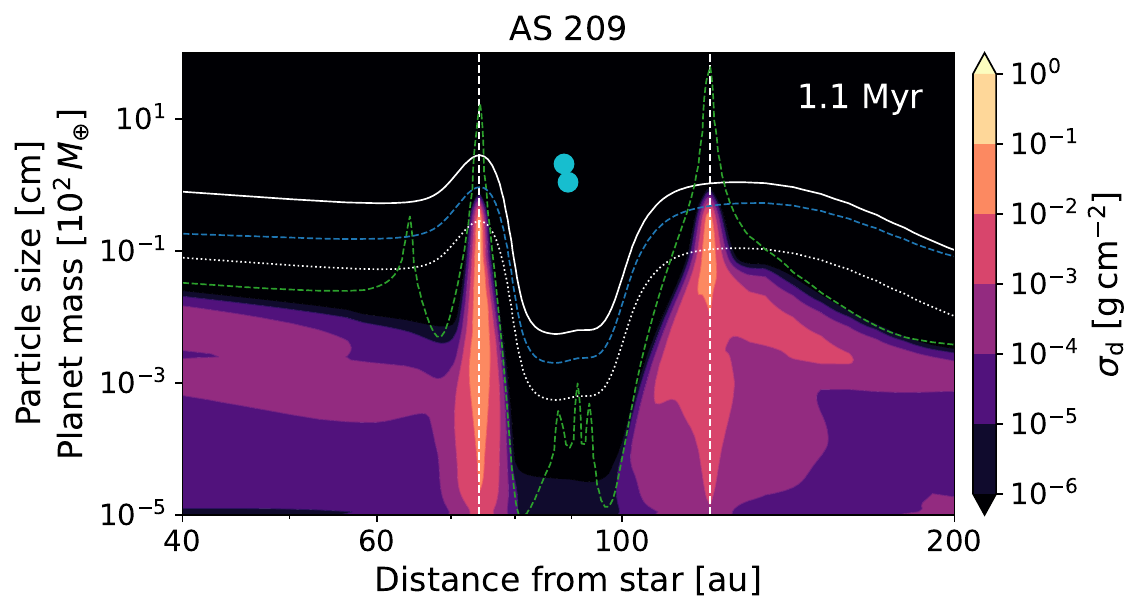}}
\resizebox{\hsize}{!}{\includegraphics{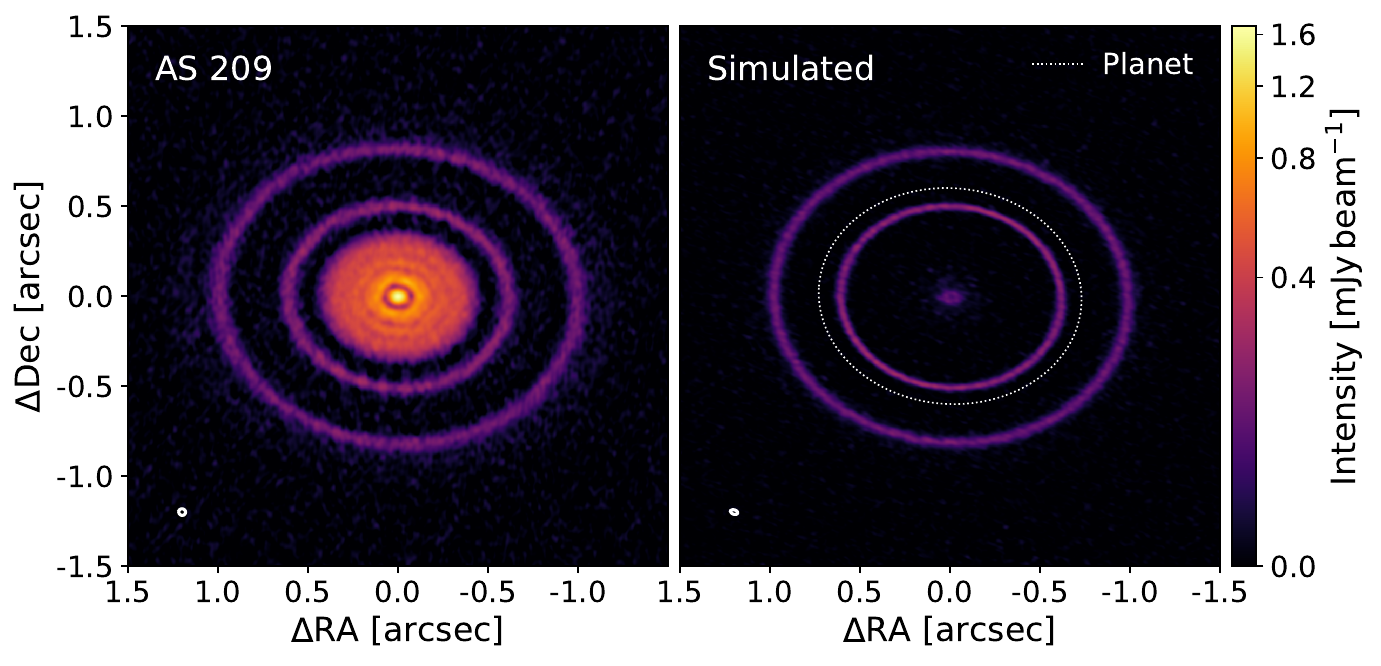}}
\caption{Dust distribution (top) and synthetic observation (bottom right) at 1.1 Myr, compared to AS 209 (bottom left; \citealp{Andrews18,Guzman18}). Vertical lines indicate the target rings. The white dashed ellipse shows the orbit of the most massive giant. Other notation follows Fig. \ref{fig:fidu_obs1}.}
\label{fig:AS209}
\end{figure}

AS 209 features rings at 74 and 120 au. Its central star ($1.2\,M_{\odot}$, $T_{\rm eff} = 4266$ K, $R_{\star} = 2.2\,R_{\odot}$) was modeled with $r_{\rm infall} = 130\,{\rm au}$ and a fragmentation velocity of $250\,{\rm cm\,s^{-1}}$.

At 1.1 Myr after the onset of late infall (Fig. \ref{fig:AS209}, top), two giant planets (207 and $109\,M_{\oplus}$) at 89 au are locked in co-orbital resonance. Their combined torques have carved a broad, deep gap, halting their migration and creating pressure bumps that align with the observed B74 and B120 rings. Planetesimal formation will begin in the inner ring shortly hereafter (1.2 Myr).

The synthetic observation (Fig. \ref{fig:AS209}, bottom) utilizes the template of Elias 24, adjusted for distance and geometry. The simulated B120 ring ($15.2\,M_{\oplus}$) reaches $\tau_{\rm max} = 0.41$ ($0.14\,{\rm Jy\, arcsec^{-2}}$), making it slightly brighter and more optically thick than the observation ($\tau_{\rm max} = 0.32$). Similarly, the simulated B74 ring ($12.4\,M_{\oplus}$) yields $\tau_{\rm max} = 0.40$, exceeding the observed $\tau_{\rm max} = 0.27$. Their intensities in $\rm mJy\, beam^{-1}$ are slightly lower than the observed image because the Elias 24 template has a different beam size than the DSHARP observation of AS 209. As with HD 163296, the simulated rings are narrower than the observations, and the inner disk is depleted. Notably, simulations producing only a single giant planet typically fail to replicate the wide spacing between these two rings.

In summary, our late-infall model successfully replicates the radial positions, intensities, and optical depths of representative DSHARP rings within acceptable limits. The lower dust mass requirements in our rings compared to \cite{Dullemond18} stem from differing opacity assumptions. While current observations of HD 163296 and AS 209 have not yet confirmed the presence of late-infall streamers, the coexistence of multiple rings and infall signatures in sources like HL Tau \citep{HLTau,Yen19,Garufi22} and GM Aur \citep{Huang20,Huang21} suggests that this mechanism is a viable pathway for disk evolution.

\section{Discussions} \label{sec:discuss}

\subsection{Planet and disk evolution} \label{ssec:dis_pebble}

\subsubsection{Pebble accretion}

The formation of multigenerational planetesimals and multiple ring substructures in our model hinges on the rapid growth of giant planets. A critical prerequisite for this growth is efficient pebble accretion. In a smooth disk, a protoplanet typically requires a mass exceeding that of the Moon to enter the efficient pebble accretion regime. This threshold is far above the initial planetesimal masses in our model. However, within the infall-induced local pressure maximum, the sub/super-Keplerian gas speed, $\Delta v = \eta v_{\rm K}$, is significantly reduced. This reduction expands the Bondi accretion radius and lowers the transition mass between the Bondi and Hill regimes. The pebble accretion onset mass, accounting for planetary eccentricity, is expressed as \citep[][modified by \citealp{Lau22}]{Visser16, Ormel17}:
\begin{equation}
    M_{\rm PA,onset} = \text{St}[\max (|\eta|,0.76e)]^3 M_{\star}.
\end{equation}
We evaluated this condition for the earliest-maturing giant in Fiducial Simulation 1. At its birth ($t = 0.245\,{\rm Myr}$, $r = 71.5\,{\rm au}$), the pressure gradient parameter was as low as $|\eta| = 3 \times 10^{-5}$, which stood in stark contrast to the $|\eta| = 0.01$ found in a typical smooth disk. This yielded an onset mass of $M_{\rm PA,onset} = 2 \times 10^{-9}\, M_{\oplus}$, which is well below the planetesimal's initial mass of $3 \times 10^{-4}\, M_{\oplus}$. From $t \geq 0.25\,{\rm Myr}$, mutual scattering increased the eccentricity ($e > 2 \times 10^{-3}$), which subsequently dominated the onset mass calculation and raised $M_{\rm PA,onset}$ above $2 \times 10^{-4}\, M_{\oplus}$. Nevertheless, the planet's mass consistently stayed above this moving threshold, allowing for uninterrupted growth. For unsuccessful planetesimals, their masses fall below $M_{\rm PA,onset}$ as scattering-induced eccentricity rises, explaining the ``oligarchic'' growth pattern observed in our results. Furthermore, while the ``loose Bondi regime'' \citep{Ormel10} can suppress the accretion of weakly-coupled particles when their stopping time ($t_{\rm stop} = \text{St}/\Omega_{\rm K}$) exceeds the passage time ($t_{\rm pass} = Gm_{\rm p}/(\Delta v+\Omega_{\rm K}R_{\rm H})^3$ with the planet's Hill radius $R_{\rm H}$), the low relative velocity $\Delta v$ in our pressure bumps sufficiently prolongs $t_{\rm pass}$. Under these conditions, the regime only applies for ${\rm St} \gtrsim 10$, far exceeding the fragmentation-limited values (${\rm St} < 1$) in our disk. Thus, large particles remain available for efficient accretion.

\subsubsection{Gas accretion}

The gas accretion rate during the runaway phase is determined by the minimum of three values: the Kelvin-Helmholtz contraction rate (Eq. \ref{eq:KH}), the local gas supply rate (Eq. \ref{eq:gas_flow}), and the global disk accretion rate (Eq. \ref{eq:gas_acc}). In our simulations, the global disk accretion rate typically serves as the limiting constraint due to the low viscosity we assumed. In contrast, the Kelvin-Helmholtz contraction timescale is consistently so short that the corresponding accretion rate is unrealistically high, and thus it never becomes the limiting factor in the runaway phase. While our prescriptions neglect the impact of eccentricity on gas accretion, recent studies suggest that moderate eccentricities may actually enhance accretion rates \citep{Bailey21, LiChen23}. Furthermore, our choice of a relatively low envelope opacity ($\kappa = 0.01\,{\rm cm^2\,g^{-1}}$) allows runaway accretion to trigger early. Consequently, in our simulations, massive embryos scattered to high-eccentricity orbits ($e \gtrsim 0.1$) can continue to grow via gas accretion even after pebble accretion is halted, eventually leading to eccentricity damping. These planets typically possess low core masses (a few $M_{\oplus}$), lower than current estimates for Solar System giants.

\subsubsection{Multigenerational planetesimal formation}

Historically, planet-induced pressure bumps were assumed to form primarily at the outer edge of planetary gaps. However, hydrodynamic simulations have demonstrated that bumps can emerge on the inner side as well \citep[e.g.,][]{Malik15, Dipierro16, Lega25}. In our model, these inner bumps are the primary sites for second-generation planetesimal formation. This is achieved by explicitly modeling planetary torques for gap opening rather than local $\alpha$-modifications, aided by the low viscosity and infall-induced density enhancements. The persistence of these bumps is maintained by the planet’s inward migration. For Fiducial Simulation 1 at 0.75 Myr, we found the inner pressure bump width to be $\sigma = 1.9\,{\rm au}$ via Gaussian fitting. The viscous dissipation timescale for such a bump was $t_{\rm visc} = \sigma^2/\nu \approx 1.6 \times 10^5\,{\rm yr}$. In contrast, the time required for the planet (at $a = 48.7\,{\rm au}$) to migrate across this width was $t_{\rm migr} = \tau_{a}(\sigma/a) \approx 2.9 \times 10^4\,{\rm yr}$. Since $t_{\rm migr} \ll t_{\rm visc}$, the migrating planet effectively provides continuous dynamical support to the bump, allowing it to ``sweep'' across the disk and produce a widely distributed population of planetesimals over an extended duration.

Our simulations were terminated at 2 Myr, prior to the onset of disk dispersal via internal photoevaporation. However, multiple studies \citep[e.g.,][]{Lau25,Ying26} have demonstrated that strong dust traps can also form during this late stage as gas gaps or cavities open. This mechanism could introduce an even younger generation of planetesimals to the system, occurring millions of years after the end of our current simulations.

We note that the present model does not include mass loss or surface erosion of planetesimals after their formation. In particular, planetesimals scattered from beyond the major ice lines to the inner disk may experience volatile sublimation or thermal surface ablation as they are exposed to higher temperatures. Such processes have been investigated in multiple studies \citep[e.g.,][]{DAngelo15,Eriksson21,RibeiroDeSousa24} and could modify the masses and compositions of inward-scattered planetesimals. In addition, we do not consider erosion by the disk gas, which can become efficient for planetesimals on highly eccentric orbits owing to their large relative velocities with respect to the gas \citep{Rozner20,Cedenblad21,Schonau23}. Nevertheless, this process is not expected to affect the growth of the giant planets formed in our simulations, as their progenitor planetesimals remain on nearly circular orbits throughout their growth. Assessing these effects is beyond the scope of the present work.

\subsection{Other potential effects of infall} \label{ssec:dis_infall}

In this work, we developed a simplified model for infall-induced dust traps by considering only local density enhancements. While our modeling framework, \texttt{DustPy}, facilitates a realistic treatment of dust coagulation and multi-species dynamics, its dimensionality inherently limits the complexity of the gas dynamics it can resolve. In contrast, 2D and 3D hydrodynamic simulations of disks perturbed by late infall have revealed a diverse range of non-axisymmetric structures. For instance, late infall can trigger the RWI near the centrifugal radius, generating long-lived vortices that trap dust even more efficiently than axisymmetric rings \citep{Bae15, Kuznetsova22}, further facilitating planetesimal formation. Furthermore, if the infalling material arrives with a significant angular momentum mismatch or high inclination relative to the disk plane, it can drive the formation of warps or even secondary misaligned disks \citep{Kuffmeier21}. In regimes of high mass-loading, infall may even trigger gravitational instability, potentially leading to the direct fragmentation of the disk and the rapid formation of giant planets \citep{Speedie25}.

Conversely, the impact of infall is not universally favorable for planet formation. The interaction between infalling streamers and the disk can generate significant turbulence and localized shock heating \citep{Huang23}. As demonstrated in our parameter study (Section \ref{ssec:res_param}; see also \citealp{Zhao25}), enhanced turbulence (represented by a higher viscosity $\alpha$) can be detrimental to the survival of pressure bumps, potentially smoothing out the substructures required for dust accumulation and the subsequent formation of massive planets. Additionally, the thermal energy released by infall-induced shocks can sublimate volatile species, such as SO, fundamentally altering the disk's thermal and chemical landscape \citep{Shariff22}. Future modeling efforts should therefore incorporate these three-dimensional and thermochemical effects to fully capture the chaotic, yet replenishing, environment of disks subject to late-stage infall.

In this work, we assumed that late infall began at $t=0$, when the dust in the disk was initialized as sub-micron-sized grains. In reality, late infall is expected to occur after some period of disk evolution, during which the primordial dust would have already undergone substantial coagulation. We investigated the effect of delayed infall in \cite{Zhao25} by initiating infall at $t=50$ kyr and 100 kyr. A later onset changed the amount and size distribution of the primordial dust available to be trapped in the pressure bump, but the resulting planetesimal mass differed only slightly from the fiducial case. This is because planetesimal formation in the infall-induced pressure bump is dominated by the dust supplied by the late infall itself, while the primordial disk dust makes only a secondary contribution. We therefore do not expect our assumption of initiating late infall at $t=0$ to qualitatively affect the conclusions of this work.

\subsection{Detectability of planetary bodies} \label{ssec:dis_exoplanet}

\begin{figure}[htp]
\resizebox{\hsize}{!}{\includegraphics{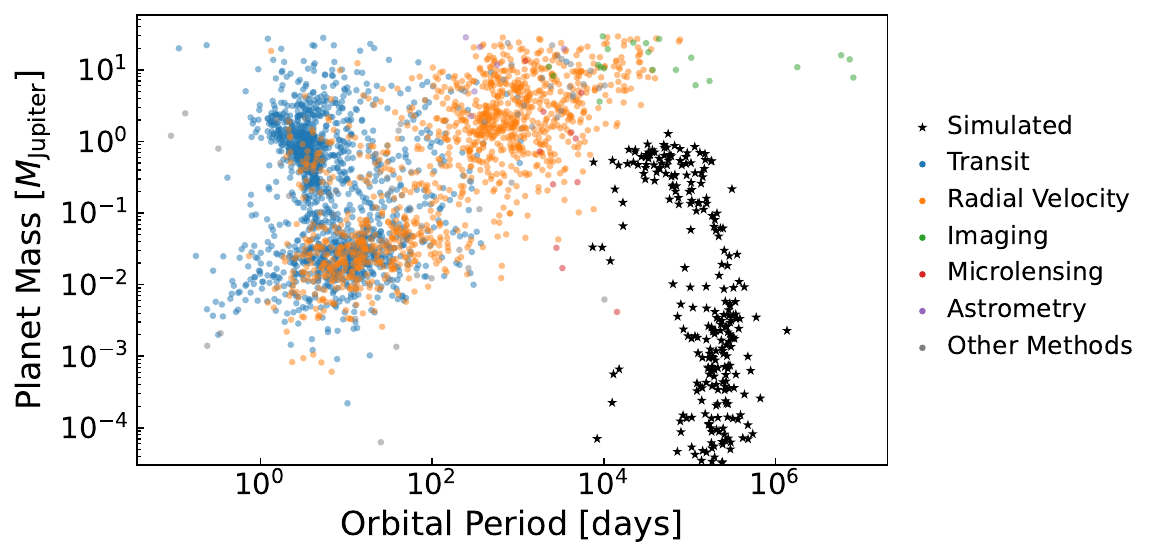}}
\caption{Distribution of orbital period and planet mass for confirmed exoplanets (circles) and the final simulated planets from all parameter setups (black stars). Confirmed exoplanet data are retrieved from the NASA Exoplanet Archive \citep{Christiansen25}, current as of May 21, 2026. Circle colors denote the detection methods, where ``Other Methods'' include transit timing variations, eclipse timing variations, pulsation timing variations, pulsar timing, and orbital brightness modulation.}
\label{fig:exoplanet}
\end{figure}

To date, over 6,000 exoplanets have been confirmed. Fig. \ref{fig:exoplanet} displays the population with known masses and orbital periods (circles), with colors distinguishing the different detection methods. The primary techniques for detecting long-period, massive planets are radial velocity, microlensing, and imaging. While radial velocity and microlensing can detect lower-mass bodies down to Earth-mass limits, they are generally insensitive to extremely long orbital periods ($> 10^5$ days). On the contrary, imaging excels at detecting planets with extremely long periods ($> 10^6$ days) but is currently restricted to relatively young, high-mass objects, typically several times the mass of Jupiter. Furthermore, imaging can identify accreting protoplanets still embedded in their natal disks, such as those around PDS 70 \citep{Keppler18} and WISPIT 2 \citep{vanCapelleveen25}. In Fig. \ref{fig:exoplanet}, black stars denote the final planetary populations produced across all of our parameter setups in Fig. \ref{fig:param}. Unfortunately, nearly all planets fall outside the current observational threshold due to insufficient mass and excessively long orbital periods. The candidate closest to the detectable regime is the innermost giant planet of Simulation 5 in the inward-shifted $r_{\rm infall}$ setup, which possesses an orbital period of 7,572 days and a mass of $0.5\,M_{\rm Jup}$. This highlights a significant observational gap: the planetary architectures resulting from late-stage infall onto the outer disk are robust in terms of formation physics but remain largely elusive to current exoplanet surveys. Although Gaia astrometry is expected to become a powerful method for detecting exoplanets, the masses and orbital periods of the planets produced in our simulations remain outside Gaia's detection limits \citep{Lammers26}. Nevertheless, molecular line observations offer a promising way to unveil these moderately-massive embedded planets. \cite{Yoshida26} recently observed silicon sulfide isotopologues in the PDS 66 disk, finding evidence for an accreting super-Earth ($<10\,M_{\oplus}$) at 60 au. The properties of this candidate align closely with the planets formed in our models, demonstrating the possible detectability of our simulated population during the gas disk phase.

The abundant population of small bodies formed in our simulations, while remaining far below the detection thresholds of current exoplanet surveys, could produce observable dust via collisional grinding during the debris disk phase following gas dispersal. Recent observations from the ARKS survey \citep{Marino26,Han26}, which characterized the radial dust distribution across 24 debris disks, provide a direct proxy for the underlying spatial configuration of parent planetesimals. In the ARKS sample, dust rings tracing distinct planetesimal belts are ubiquitously detected at heliocentric distances of tens to hundreds of au. While our models typically yield broadly distributed planetesimal populations, localized clustering into narrow belts does occur under certain conditions (e.g., the G3 planetesimals in Fiducial Simulation 2). This variability may result in detectable dust substructures in the later debris phase, a possibility that warrants future investigation.

\subsection{Comparisons to the Solar System} \label{ssec:dis_kuiper}

Small bodies in the Kuiper belt are traditionally categorized into three distinct dynamical populations: a scattered group (high eccentricity and inclination), a resonant group (captured in mean-motion resonances with Neptune), and a dynamically cold group (unperturbed by massive planets). \citet{Lau25} proposed that these populations could be formed via disk dissipation, where internal photoevaporation drives a planetesimal-forming dust trap outward following the birth of giant planets. In their model, the scattered and resonant groups form during the outward migration of the outermost giant, whereas the dynamically cold group forms later in the outer disk after migration has stalled.

Our model similarly produces multiple dynamical classes of small bodies in the outer disk, but through a different mechanism. In Fiducial Simulation 2, the G1 and G2 planetesimals scattered outward by migrating giants constitute a hot, scattered population ($e \gtrsim 0.1$), whereas the less strongly perturbed G3 planetesimals form a resonant population concentrated at the 5:3 and 2:1 period ratios. This capacity of our model to produce distinct dynamical components also aligns well with exo-Kuiper belt observations. The vertical structures of debris disks analyzed in the ARKS survey point to the presence of multiple, coexisting dynamical populations \citep{Zawadzki26}.

A similar coexistence of resonant and non-resonant architectures can be found in the inner regions of planetary systems. In the main belt of the Solar System, Jupiter’s gravity stabilizes specific resonant groups, such as the Hilda asteroids, while clearing out unstable regions to create the Kirkwood gaps. This inner-system dynamic is successfully captured in both our Fiducial Simulations 1 and 2, where a significant population of planetesimals in the inner disk are trapped into mean-motion resonances with the innermost giant planet.

Furthermore, the isolation of the inner and outer Solar System by Jupiter is the primary explanation for the isotopic dichotomy between non-carbonaceous (NC) and carbonaceous (CC) chondrites \citep{Kruijer17, Kleine20}. This theory requires Jupiter to form within $\sim 1\,{\rm Myr}$ to establish a transport barrier for dust early enough to maintain distinct reservoirs. In our model, giant planets consistently mature within $1\,{\rm Myr}$ and carve deep gaps that isolate the inner and outer disk, providing a natural mechanism for such a dichotomy. 

\citet{Hellmann23} revealed that the abundances of different dust components within CC chondrites varied significantly over the duration of planetesimal formation. \citet{Gurrutxaga26} demonstrated that the compositional variation can be explained by the varying dust filtering and delivery rates of distinct solid components to a long-lived pressure bump, likely induced by Jupiter. Because planetesimal formation in our outermost planet-induced bumps can be long-lasting (extending beyond 2 Myr in Fiducial Simulations 2 and 5), these populations would likely exhibit similar compositional variations over time if the individual dust components were traced, mirroring the delivery trends observed in the Solar System's meteoritic record. In addition, a recent cosmochemical analysis of CC chondrites by \cite{vanKooten26} reveals the mixing of two distinct dust reservoirs, which may signify late infall replenishing the outer Solar System.

\subsection{Other recent works modeling multigenerational planet formation}

\cite{Lau24} developed a model of sequential giant planet formation triggered by a disk substructure at $\sim 5$ au. In their model, the formation of a Jupiter-mass gas giant inside a pressure bump creates a new planetesimal formation site exterior to its orbit, eventually producing a compact chain of gas and ice giants. Because these planets open gaps rapidly in the inner region (at a few au), their inward migration is limited, causing the planetesimal formation sites to move from the inside out. In contrast, the slower gap opening in the outer disk in our work allows the giant planets to migrate continuously at the Type-I speed, causing the planetesimal formation sites to move from the outside in. Furthermore, the migrating first-generation giants excite the eccentricities of the later-generation planetesimals, preventing them from growing into new giant planets. Therefore, while the growth of giant planets in our model is still successive, all the giants ultimately originate from the first-generation planetesimals.

\cite{Sandor24} modeled planet formation initiated by a transient dust trap at the dead zone boundary, created by a jump in the disk viscosity ($\alpha$). They investigated how a planet growing inside this trap affects the surrounding disk. In their model, the planet opens a gap and induces a new dust trap at its outer edge. As the planet migrates inward, this trap moves with it, continuously producing planetesimals and forming a broad planetesimal belt. In our study, planet-induced dust traps can also emerge at the inner edge of planetary gaps and move with the migrating planets, because we model gap opening using self-consistent torque deposition. Furthermore, the two-population dust evolution model \citep{Birnstiel12} used by \cite{Sandor24} has known limitations when dealing with disk substructures. It tends to overestimate dust concentration in pressure bumps \citep[see][for details]{Pfeil24}, which may have led to excessive planetesimal production in their simulations.

The concept of ``sandwiched planet formation'' was proposed by \cite{Pritchard24}, who studied dust concentration between the orbits of two massive planets. They found that the outer planet reduces the amount of dust collected outside the inner planet's orbit. This potentially allows a smaller planet to form between the two massive giants, offering an explanation for the architectures of certain exoplanetary systems. However, their study was qualitative and did not track the actual conversion of dust into physical planets, thereby neglecting planetary dynamics and its feedback on growth. In our work, the planetesimals that form in this sandwiched scenario (e.g., the G3 and G4 populations in Fiducial Simulation 1) are rapidly scattered to high eccentricities by the neighboring giant planets shortly after their formation. This dynamic instability calls into question whether a smaller, detectable planet can successfully form between two preexisting giants.

\section{Conclusions} \label{sec:conclu}

The evolution of protoplanetary disks and the efficiency of planet formation are strongly influenced by their external environment. Late infall not only replenishes the solid reservoir of a disk but also fundamentally reshapes its pressure profile. Following the work of \citet{Zhao25}, we have investigated the subsequent formation of giant planets and multigenerational planetesimals in a disk fed and perturbed by late infall. Our key findings are summarized as follows:

\begin{itemize}
    \item The infall-induced pressure bump serves as both a high-density dust reservoir and a planet migration trap. This combination keeps embryos in a pebble-rich environment, allowing the growth of gas giants at wide orbits ($\sim 70$ au) to be completed within 1 Myr. (Figs. \ref{fig:fidu1} and \ref{fig:fidu2}.)

    \item The feedback from these giant planets transforms the initial infall-induced pressure bump into multiple planet-induced bumps. These secondary traps initiate multigenerational planetesimal formation, leading to diverse planetary system architectures. (Figs. \ref{fig:fidu1}--\ref{fig:history}.)

    \item We classify planet-induced planetesimal formation into three regimes: (i) the inner-side scenario, maintained by inward planet migration and gap widening due to planetary growth; (ii) the sandwiched scenario, occurring when differential planet migration speeds concentrate dust between two giants; and (iii) the outer-side scenario, triggered by the accumulation of residual outer-disk dust. (See Section \ref{ssec:res_fidu} for detailed descriptions.)

    \item The resulting planetesimals populate multiple dynamical groups that are analogous to the distribution of small bodies in the Solar System. These dynamical states are intrinsically linked to the timing and location of planetesimal formation relative to the growth and migration of the giants. Furthermore, the rapid formation of these giants establishes a dust transport barrier, potentially leaving cosmochemical imprints similar to those observed in the Solar System. (Section \ref{ssec:dis_kuiper}.)

    \item Giant planet growth, the occurrence of planetesimal formation, and the final architecture of the planetary system all depend critically on a combination of dust, gas, and global disk properties. (Section \ref{ssec:res_param}.)

    \item Synthetic continuum images reveal that infall- and planet-induced rings resemble the multi-ring disks observed in the ALMA DSHARP survey. The planetesimal-forming rings reach optical depths consistent with observed bright rings. Our model successfully reproduces the radial locations and intensities of rings in HD 163296 and AS 209. (Figs. \ref{fig:fidu_obs1}--\ref{fig:AS209}.)
\end{itemize}

While this work focuses on axisymmetric dust traps, late infall is known to trigger more complex 3D phenomena, including spirals, vortices, warps, and gravitational instabilities. Although these high-dimensional dynamics are beyond the scope of the current 1D dust coagulation and transport model, the impact of late-stage infall on planet formation remains a vital frontier for understanding the origin and diversity of planetary systems.

\begin{acknowledgements}
JD was funded by the European Union under the European Union’s Horizon Europe Research \& Innovation Programme 101040037 (PLANETOIDS). Views and opinions expressed are however those of the author only and do not necessarily reflect those of the European Union or the European Research Council. Neither the European Union nor the granting authority can be held responsible for them. This research was supported by the Munich Institute for Astro-, Particle and BioPhysics (MIAPbP) which is funded by the Deutsche Forschungsgemeinschaft (DFG, German Research Foundation) under Germany's Excellence Strategy – EXC-2094 – 390783311.
\end{acknowledgements}

\bibliographystyle{aa}
\bibliography{ref}

\end{document}